\documentclass[12pt, a4paper]{iopart}
\usepackage{epsfig, enumerate}
\usepackage{graphicx,epsf}
\usepackage{color}
\usepackage{bm}
\usepackage{iopams}
\eqnobysec
\def\be{\begin{equation}}
\def\ee{\end{equation}}
\def\beb{\begin{equation*}}
\def\eeb{\end{equation*}}
\def\bea{\begin{eqnarray}}
\def\eea{\end{eqnarray}}
\def\beab{\begin{eqnarray*}}
\def\eeab{\end{eqnarray*}}
\def\nn{\nonumber}

\def\p{\partial}

\def\X{{{\mathbb{X}}}}
\def\XB{{{\mathbb{X}}_{\rm B}}}
\def\H{{H}}

\def\cs2{c_{\rm{s}}^2}

\def \eps {\epsilon}
\def \beg {\begin{enumerate}}
\def \en {\end{enumerate}}

\def\rhob{\rho_0}

\def\cs{c_{\rm{s}}^2}
\newcommand\eq[1]{Eq.~(\ref{#1})}

\def\fg{{\rm{flat}}}
\def\lg{{\ell}}

\def\wt{\widetilde}
\def\ipsi{{\psi_1}}

\def\iphi{{\phi_1}}
\def\iiphi{{\phi_2}}

\begin{document}

\title{Comparing different formulations of non-linear cosmological perturbation theory}
\author{Adam J.~Christopherson$^1$, Karim A.~Malik$^1$, David~R.~Matravers$^2$
and Kouji Nakamura$^3$}
\address{\vspace{2mm}$^1$Astronomy Unit,
School of Mathematical Sciences, \\
Queen Mary University of London, \\
Mile End Road, London, E1 4NS, United Kingdom.\\[2mm]
$^2$Institute of Cosmology and Gravitation,\\
  University of Portsmouth,\\
  Dennis Sciama Building, Portsmouth, PO1 3FX, United Kingdom\\[2mm]
 $^3$Optical and Infrared Astronomy Division,\\
  National Astronomical Observatory of Japan,\\ 
  Osawa 2-21-1, Mitaka, Tokyo 181-8588, Japan.}
  \eads{\mailto{a.christopherson@qmul.ac.uk}, \mailto{kouji.nakamura@nao.ac.jp}}

\begin{abstract}
We compare and contrast two different metric based formulations
of non-linear cosmological perturbation theory: the MW2009
approach in [K.~A.~Malik and D.~Wands, Phys.\ Rept.\  {\bf 475}
(2009), 1.] following Bardeen and the recent approach of the
paper KN2010 [K.~Nakamura, Advances in Astronomy {\bf 2010}
(2010), 576273].
We present each formulation separately. 
In the MW2009 approach, one considers the gauge transformations
of perturbative quantities, choosing a gauge by requiring that
certain quantities vanish, rendering all other variables gauge
invariant. 
In the KN2010 formalism, one decomposes the metric tensor into a
gauge variant and gauge invariant part from the outset.
We compare the two approaches in both the longitudinal and
uniform curvature gauges.
In the longitudinal gauge, we find that Nakamura's gauge
invariant variables correspond exactly to those in the longitudinal gauge
(i.e., for scalar perturbations, to the Bardeen potentials), and in the uniform curvature gauge we
obtain the usual relationship between gauge invariant variables
in the flat and longitudinal gauge.
Thus, we show that these two approaches are equivalent.
\end{abstract}

\maketitle

%%%%%%%%%%%%%%%%%%%%%%%%%%%%%%%%%%%%%%%%%%%%%%%%%%%%%%%%%%%%%%%%%%%%%%
%%%%%%%%%%%%%%%%%%%%%%%%%%%%%%%%%%%%%%%%%%%%%%%%%%%%%%%%%%%%%%%%%%%%%%
\section{Introduction}
\label{sec:Introduction}
%%%%%%%%%%%%%%%%%%%%%%%%%%%%%%%%%%%%%%%%%%%%%%%%%%%%%%%%%%%%%%%%%%%%%%
%%%%%%%%%%%%%%%%%%%%%%%%%%%%%%%%%%%%%%%%%%%%%%%%%%%%%%%%%%%%%%%%%%%%%%

%****************************************************************

Many problems in Physics and Applied Mathematics can be
described by non-linear systems of evolution equations.
These are notoriously difficult to solve exactly because of the
non-linearity.
An example of such a theory is General Relativity.
Einstein's equations are highly non-linear and can only be solved
exactly in a small number of useful cases.
To go beyond these solutions perturbative methods are
used.
Given a solution to the equations in the form of a metric
$g_{\mu \nu}^{(0)}$ we assume that we can approximate a
neighbouring, more general, solution $g_{\mu\nu}$ using a power
series.
Thus we express the more general solution in the form
\be
\label{app1}
g_{\mu \nu} 
= g_{\mu \nu}^{(0)} + g_{\mu \nu}^{(1)} + \frac{1}{2} g_{\mu \nu}^{(2)} + \ldots \,.
\ee
The metric $g_{\mu \nu}^{(0)}$ is called the \emph{background} and the
remaining terms are \emph{perturbations} of the background. The first
order part is
\be
g_{\mu \nu} - g_{\mu \nu}^{(0)} \simeq g_{\mu \nu}^{(1)}\,,
\ee
where the remaining terms are assumed to be negligible compared to
$g_{\mu \nu}^{(1)}$ and they are neglected at {\em first order}. In
a similar way the higher-order perturbations can be identified, so
at second order we have
\be
g_{\mu \nu} - g_{\mu \nu}^{(0)} - g_{\mu \nu}^{(1)} \simeq g_{\mu \nu}^{(2)} \,,
\ee
and so on.
This can be described simply if we introduce an infinitesimal parameter $\epsilon\ll 1$ for
the perturbation and assume that the
series can be written as
\be
g_{\mu \nu} = g_{\mu \nu}^{(0)} + \eps \bar{g}_{\mu \nu}^{(1)}
+ \frac{1}{2}\eps^{2} \bar{g}_{\mu \nu}^{(2)} + \ldots \,,
\ee
where the quantities with bars have absolute magnitudes less
than one.
In this format the orders correspond to the powers of $\eps$.
In practise it is often a nuisance to introduce the parameter
$\eps$ so  we will use the form (\ref{app1}) if there is no
confusion.
Issues of convergence can, in general, be removed by working in
a small enough neighbourhood but they should not be ignored.

%****************************************************************

Having set up the approximation (\ref{app1}) we have to substitute it
into the Einstein equations
\be
\label{Ein}
G_{\mu \nu} + \Lambda g_{\mu \nu} = T_{\mu \nu}\,,
\ee
to obtain solutions of the required order of approximation.
This is
more complicated than might be expected. Perturbations of the metric
imply perturbations of the energy momentum tensor and vice versa but
more significantly, calculation of the connection coefficients and the
Ricci tensor involve raising and lowering indices and so introduces
more terms and potentially couple terms of different orders. At zeroth
and first order this is not a problem but at higher orders it makes
calculations much more complicated. 
Even at second order there are ``proper'' second order terms,
for example $\bar{g}_{\mu \nu}^{(2)}$, and terms quadratic in
the first order quantities, for example $\bar{g}_{\mu \nu}^{(1)}$.

%****************************************************************

In addition to the complexity of the system of equations, the
split of the metric and matter variables into a background and
perturbations introduces spurious coordinate artefacts or gauge
modes as described in detail in numerous
reviews~\cite{Bardeen-1980, Kodama-Sasaki-1984, 
Mukhanov-Feldman-Brandenberger-1992, Malik:2008im, Nakamura:2010yg}. 
We only give a brief explanation, here.

%****************************************************************

As we are interested in cosmology, we will assume that the real
universe is described by a perturbed Friedmann-Robertson-Walker
(FRW) metric and that the background, or unperturbed spacetime,
is described by an exact FRW metric.
So we have in effect two spacetimes -- one ``physical''
and one ``fictitious''.
We label points in the background by coordinates $\{ x^{\mu} \}$
and a one-to-one-map between points in the background and points
in the physical spacetime maps these coordinates from the
background to the physical spacetime.
We refer to this one-to-one-map as a gauge choice.
A change in the map is called a gauge transformation and
this may be carried out in a number of ways, see for instance
the recent reviews ~\cite{Malik:2008yp,Malik:2008im,Nakamura:2010yg}.
A perturbation of some quantity is the difference between the
value at a point in the physical spacetime and the value
at the point in the background with the same coordinates.
Clearly such a perturbation depends on the above gauge choice.

%****************************************************************

It is important to note that a gauge transformation is different
to a coordinate transformation which changes the labels on
points in the physical and background spacetimes together, and
so it does not change the gauge.
A simple example of a gauge transformation is an implementation
of a coordinate change in the physical spacetime but not in the
background.
This changes the correspondence between the points in the two
spacetimes, so it is a gauge transformation.
It is easy to see that a scalar, e.g. the energy density $\rho$,
which is (at least) time dependent, will not be invariant under
such a transformation.
Furthermore if a gauge is chosen to simplify the metric on the
physical spacetime and some residual gauge freedom remains then
spurious gauge mode solutions may appear.
For these reasons gauge invariant formulations of perturbations
and other special gauge fixing forms have been sought.
These fall into two broad classes, (1) those
following the general pattern of
Bardeen~\cite{Stewart:1974uz,Bardeen-1980,Kodama-Sasaki-1984,Mukhanov-Feldman-Brandenberger-1992,Malik:2008im,Nakamura:2010yg}
and (2) those following the covariant form developed by Ellis
and Bruni ~\cite{Ellis:1989jt,Ellis:1989ju,Langlois:2005ii,Enqvist:2006fs}.

%****************************************************************

At first order in the perturbations the approaches in class (1) above
differ largely due to different splits of the spatial part of
the metric, and notation and sign conventions.
At second order things are more complicated and there are at
least two different approaches. Relating the 
approach used by Malik and Wands ~\cite{Malik:2008im} and the 
Nakamura approach ~\cite{Nakamura:2010yg} (hereafter referred to as MW2009 and KN2010, respectively)
 is the aim of this paper.
We aim at highlighting similarities and differences of the two
subclasses of approaches following Bardeen and try to keep the
mathematical background discussions to a minimum, referring the 
interested reader to the reviews listed above and the original
literature cited therein.
However, in order to introduce some of the quantities and
concepts used later on, we here briefly review the formulation
of perturbation theory in a more rigorous sense.

%****************************************************************

In relativistic perturbation theory, we
consider two distinct spacetimes: the `background' spacetime,
denoted ${\cal{M}}_0$ and the `physical' spacetime denoted
${\cal{M}}$, following Stewart ~\cite{stewart}. 
The physical manifold is nature itself, and we want to describe
the properties of this spacetime through the perturbations. 
On the other hand, the background spacetime is a mere reference
spacetime on which to calculate perturbations.
We then introduce a point identification map
${\cal{X}}:{\cal{M}}_0\to{\cal{M}}$ relating points on the
unperturbed manifold to points on the perturbed manifold.
By virtue of the pullback of this point identification map
(denoted ${\cal{X}}^*$), we may treat a tensor field on the
physical spacetime ${\bf T}$ as a tensor field on the background
spacetime ${\cal{X}}^*{\bf T}$, which we will often denote,
equivalently, as ${}_{\cal{X}}{\bf T}$. 
The choice of map ${\cal{X}}$, the
{\it gauge choice}, is not unique; if we choose another
different map, ${\cal{Y}}$, the pulled back variables are then
different representations of the tensor ${\bf T}$.
These two gauge choices induce a diffeomorphism 
$\Phi: {\cal{M}}_0 \to {\cal{M}}_0$, defined as 
$\Phi:={\cal{X}}^{-1}\circ{\cal{Y}}$.

%****************************************************************

We conclude this introduction by presenting
some essential equations from the reviews
~\cite{Malik:2008yp,Malik:2008im}, in the notation which is 
used in the standard literature 
(e.g.~ Refs.~\cite{Bardeen-1980, Kodama-Sasaki-1984, Mukhanov-Feldman-Brandenberger-1992}).
As in the case of the metric tensor \eq{app1} above, we assume
that any tensorial quantity ${\bf{T}}$ can be expanded into a
power series by
\bea
\label{expandT}
{\bf{T}} &=& {\bf{T}}_{0} + \delta {\bf{T}} \nonumber\,,\\
\delta {\bf{T}} &=& {\bf{T}}_{1} + \frac{1}{2} {\bf{T}}_{2} 
+ \frac{1}{3!}{\bf{T}}_{3} + \ldots\,,
\eea
where the subscripts denote the order of the perturbation. 
The change in a perturbed quantity (to a certain order) induced
by a gauge transformation is given by the exponential map, once
the generating vector of the gauge transformation, $\xi^\mu$,
has been specified.
The exponential map is 
\be
\label{Ttransgeneral}
\Phi^*{\bf T}\equiv\widetilde{\bf T}=e^{\pounds_{\xi}}{\bf T}\,,
\ee
where $\pounds_{\xi}$ denotes the Lie derivative with respect to
$\xi^\mu$.
Expanding the exponential map and using \eq{expandT}
we obtain in the background, at first order and at second order,
respectively, 
\bea
\label{transform}
\wt{{\bf T}}_{0} & = & {\bf T}_{0}\,,\nonumber \\
\wt{{\bf T}}_{1} & = & {\bf T}_{1}
+ \pounds_{\xi_{1}}{\bf T}_{0} \,,\nonumber \\
\wt{{\bf T}}_{2} & = & {\bf T}_{2}
+ \pounds_{\xi_{2}}{\bf T}_{0}
+  \pounds_{\xi_{1}}^{2}{\bf T}_{0}
+ 2 \pounds_{\xi_{1}}{\bf T}_{1}\,,
\eea
where $\xi^\mu$ is the vector field generating the transformation
and is expanded order by order as
$\xi^\mu\equiv \epsilon \xi_1^{\mu} +\frac{1}{2}\epsilon^2\xi_2^{\mu}
+O(\epsilon^3)$.
The exponential map can also be applied to the coordinates
$x^\mu$ to obtain the following relationship between coordinates
at two points, $p$ and $q$ 
\be
\label{defcoordtrans}
{x^\mu}( q)
= e^{\xi^\lambda \frac{\p}{\p x^\lambda}\big|_p} \
x^\mu( p)\,.
\ee 
Expanding to second order gives
\be
\label{eq:coords}
x^\mu(q)=x^\mu(p)+\epsilon\xi_1^\mu(p)
+\frac{1}{2}\epsilon^2\Big(\xi^\mu_{1,\lambda}(p)\xi_{1}^{\lambda}(p)+\xi_2^\mu(p)\Big)\,.
\ee
Note that this coordinate relationship is not required to
perform calculations in the active approach, but will be useful
for later discussion.

%****************************************************************

On the other hand, when evaluating the gauge transformation
rule, a different point of view is adopted in the formulation in
KN2010, where the change in a perturbed
quantity induced by a gauge transformation $\Phi$ is represented
in the general form of the Taylor expansion of $\Phi^{*}$
\begin{eqnarray}
  \label{eq:Taylor-exp-of-a-tensor-field-general}
  {\bf T}(q)
  =
  (\Phi^{*}{\bf T})(p)
  =
  {\bf T}(p)
  + \epsilon \left.{\pounds}_{\xi_{1}}{\bf T}\right|_{p}
  + \frac{1}{2} \epsilon^{2} \left.
    \left(
      {\pounds}_{\xi_{2}}
      + {\pounds}_{\xi_{1}}^{2}
    \right)
    {\bf T}
  \right|_{p}
  + O(\epsilon^{3}).
\end{eqnarray}
As shown by Bruni and 
coworkers ~\cite{M.Bruni-S.Matarrese-S.Mollerach-S.Soonego-CQG1997,S.Sonego-M.Bruni-CMP1998,Matarrese-Mollerach-Bruni-1998,Bruni-Gualtieri-Sopuerta-2003,Sopuerta-Bruni-Gualtieri-2004},
the Taylor expansion of the pull-back of tensor field is always
given in the form of Eq.~(\ref{expandT}), even if $\Phi^{*}$ is
not an exponential map. 
In this sense, the Taylor expansion
(\ref{eq:Taylor-exp-of-a-tensor-field-general}) represents the
perturbative expansion of a wider class of diffeomorphisms
than exponential maps.
Through this general formula
(\ref{eq:Taylor-exp-of-a-tensor-field-general}) and the
perturbative expansion (\ref{expandT}) of the variable
${{\bf T}}$, we can reach the same order-by-order
gauge transformation rules as Eq.~(\ref{transform}).

%****************************************************************

Although the gauge transformation rules at each order appear to
have the same form in the two different
formulations ~\cite{Malik:2008im, Nakamura:2010yg}, we should
point out that the approaches to obtain the transformations are
conceptually different.
As mentioned above, the Taylor expansion
(\ref{eq:Taylor-exp-of-a-tensor-field-general}) is 
valid for a wider class of diffeomorphisms beyond the exponential map.
Although we may regard
Eq.~(\ref{eq:Taylor-exp-of-a-tensor-field-general}) as that of
an exponential map through a special choice of 
$\xi_{2}^{\mu}$ (e.g.~Ref.~\cite{M.Bruni-S.Matarrese-S.Mollerach-S.Soonego-CQG1997}),
there is no guarantee that this is true for any choice of
$\xi_{2}^{\mu}$. 
However it seems that, when working to a particular order in
perturbation theory, one can always use an exponential map that
generates the specified gauge transformation to that order.
Therefore, while conceptually the gauge transformation rules
used by the different formulations are not the same, this is
only a philosophical issue and, in practice, the two are
equivalent.
%Thus, we can expect that these two formulations in MW2009 and
%KN2010 are conceptually different but are equivalent from
%practical point of view.

%****************************************************************

In this paper, we discuss the equivalence of these two
formulations in MW2009 and KN2010, through clarifying the
correspondence of the variables.
Having introduced the basics, we now go on to tackle the main
topic of this article, namely, obtaining the relationship
between the second-order metric perturbation in MW2009 and that
in KN2010.
It should be noted that the two formulations use very different notation.
We should note that they use very different notation in these
two formulations.
Rather than try and enforce a common notation between the two,
we present each in its conventional notation, relating them to
one another at the end.
Further, we discuss two different gauge fixing. 
One is the Poisson gauge (longitudinal gauge) choice and the
other is the flat gauge choice.
In MW2009, they showed that these two gauge fixings are complete
gauge-fixing, while there is no explicit gauge-fixing in the
formulation in KN2010.
We clarify the correspondence between the variables in MW2009
and KN2010 through these two gauge-fixing.

%****************************************************************

We also have to emphasise that although this paper is not a
complete survey of the many different formulations of
second-order cosmological perturbation theory, the
correspondence which is clarified here is a useful check of the
equivalence of different formulations and will hopefully lead to
a consensus in the community.

%****************************************************************

This paper is organised as follows.
In the next section, we describe first order perturbations, 
first in the MW2009 approach and then in the KN2010 approach. 
In
\S\ref{sec:Formulations_for_the_second-order_cosmological_perturbations}
we review  second order perturbation theory, again starting with
the MW2009 approach and then the KN2010 approach.
In \S\ref{sec:Comparison_of_different_formulations} we compare
the two approaches, in both the longitudinal (or Poisson) gauge,
and the uniform curvature gauge.
Finally, we summarise our results in
\S\ref{sec:Summary_and_discussions}.

%****************************************************************

If not otherwise stated we use conformal time,
$\eta$, related to coordinate time $t$ by $dt=a d\eta$, where
$a(\eta)$ is the scale factor, see Eq.~(\ref{eq:background-metric}), throughout.
Derivatives with respect to conformal time are denoted by a
prime, and the Hubble parameter is defined, in terms of
conformal time, as $H=a' / a$.
Greek indices, $\mu,\nu,\lambda$, run from $0\ldots 3$, and
lower case Latin indices, $i,j,k$, run from $1\ldots3$.
We also make use of abstract indices $a,b,c,$ in parts.

%****************************************************************

%%%%%%%%%%%%%%%%%%%%%%%%%%%%%%%%%%%%%%%%%%%%%%%%%%%%%%%%%%%%%%%%%%%%%%
%%%%%%%%%%%%%%%%%%%%%%%%%%%%%%%%%%%%%%%%%%%%%%%%%%%%%%%%%%%%%%%%%%%%%%
\section{First order cosmological perturbations}
\label{sec:first_order_cosmological_perturbations}
%%%%%%%%%%%%%%%%%%%%%%%%%%%%%%%%%%%%%%%%%%%%%%%%%%%%%%%%%%%%%%%%%%%%%%
%%%%%%%%%%%%%%%%%%%%%%%%%%%%%%%%%%%%%%%%%%%%%%%%%%%%%%%%%%%%%%%%%%%%%%

%****************************************************************

The background spacetime ${\cal M}_{0}$ considered in
cosmological perturbation theory is a homogeneous, isotropic
Friedmann-Robertson-Walker universe foliated by three
dimensional hypersurfaces $\Sigma(\eta)$, parametrised by
conformal time $\eta$.
In this paper we restrict ourselves to considering flat spatial
hypersurfaces, and the line element for this spacetime is
\begin{eqnarray}
  \label{eq:background-metric}
  ds^2 = a^2(\eta)\Big[-d \eta^2+\delta_{ij}dx^i dx^j\Big]\,,
\end{eqnarray}
where $a=a(\eta)$ is the scale factor and $\delta_{ij}$ is the
metric on the flat space.
The full spacetime metric is then expanded, as in Eq.~(\ref{expandT}), as
\beb 
g_{\mu \nu}=g_{\mu \nu}^{(0)}+\delta g_{\mu \nu}^{(1)}+\frac{1}{2}\delta g_{\mu\nu}^{(2)}+\cdots\,,
\eeb
in the notation of MW2009.
Alternatively, one can represent the background spacetime metric as
\be 
\label{eq:background-KN2010}
g_{ab}=a^2(\eta)\Big[-(d\eta)_a (d\eta)_b + \delta_{ij}(dx^i)_a (dx^j)_b\Big]\,,
\ee
in the abstract index notation with the full spacetime metric
then being expanded as 
\begin{eqnarray}
  \label{eq:metric-expansion}
  {g}_{ab}
  =
  g_{ab}^{(0)} + \epsilon h_{ab}
  + \frac{1}{2} \epsilon^{2} l_{ab} + O(\epsilon^{3}).
\end{eqnarray}
Equations (\ref{eq:background-KN2010}) and
(\ref{eq:metric-expansion}) are the abstract index notation in
KN2010.

%****************************************************************

The metric components at each order in perturbation theory can
then be expanded into scalar, vector and tensor components,
according to their transformation behaviour on spatial
hypersurfaces.

%****************************************************************

%%%%%%%%%%%%%%%%%%%%%%%%%%%%%%%%%%%%%%%%%%%%%%%%%%%%%%%%%%%%%%%%%%%%%%
\subsection{MW2009 formulation}
\label{sec:standard_formulation}
%%%%%%%%%%%%%%%%%%%%%%%%%%%%%%%%%%%%%%%%%%%%%%%%%%%%%%%%%%%%%%%%%%%%%%

%****************************************************************

At linear, or first, order in perturbation theory the general
scalar, vector and tensor perturbations to the flat ($K=0$) FRW
background spacetime can be expressed in the line element
\be
\label{eq:lineelement}
ds^2=a^2(\eta)\left[-(1+2\iphi)d\eta^2+2B_{1i}dx^id\eta+(\delta_{ij}+2C_{1ij})dx^idx^j\right]\,.
\ee
The perturbations of the spatial components of the metric can
then be further decomposed as ~\cite{Malik:2008im}
\begin{eqnarray}
B_{1i} &= B_{1,i}-S_{1i} \,,\\
C_{1ij} &= -\ipsi \delta_{ij}+E_{1,ij}+F_{1(i,j)}+\frac{1}{2}h_{1ij}\,,
\end{eqnarray}
where $\iphi, B_1, \ipsi$ and $E_1$ are scalar metric
perturbations, $S_{1i}$ and $F_{1i}$ are divergence-free vector
perturbations, and $h_{ij}$ is a transverse, traceless tensor
perturbation. 
In the notation of KN2010, this is then
\begin{eqnarray}
\label{eq:hetaeta}
  &&
  h_{\eta\eta} = - 2 a^{2} \phi_{1}
  \label{eq:Malik:2008im-hetaeta-phi1-def}
  , \\
  &&
  \label{eq:hieta}
  h_{i\eta} = a^{2}B_{1i} \equiv a^{2}D_{i}B_{1} - a^{2}S_{1i}
  \label{eq:Malik:2008im-hieta-B-S-def}
  , \\
  &&
  \label{eq:hij}
  h_{ij} = 2 a^{2} C_{1ij} = 2 a^{2} \left(
    - \psi_{1} \delta_{ij} + D_{i}D_{j}E_{1}
    + D_{(i}F_{1j)}
    + \frac{1}{2} h_{1ij}
  \right)\,,
  \label{eq:Malik:2008im-hij-psi1-E1-F1-h1ij-def}
 \end{eqnarray}
where $D_{i}$ is formally the covariant derivative associated
with the spatial metric $\delta_{ij}$ and, in practice for this
work, it reduces to a partial derivative denoted by a comma.
Consistently perturbing the spacetime will naturally invoke
perturbations to its matter content as well.
In the following, however, we shall only use the energy density.

%****************************************************************

Before studying the transformation behaviour of perturbations at
first order, we split the generating vector $\xi_1^\mu$ into a
scalar temporal part $\alpha_1$ and a spatial scalar and
divergence-free vector part, respectively $\beta_1$ and
$\gamma_1{}^i$, as 
\be 
\xi_1^\mu=(\alpha_1,\beta_{1,}{}^i+\gamma_1{}^i)\,.
\ee
We can then consider transformations of different types of
perturbation independently, since they decouple at linear
order. 
For example, \eq{transform} implies that the energy density
perturbation transforms, at first order, as
\be
\label{eq:rhotrans1}
\widetilde{\delta\rho_1}=\delta\rho_1+\rhob'\alpha_1\,,
\ee
where we have used the fact that the Lie derivative, when acting
on a scalar function, is just $\pounds_{\xi}=\xi^\mu(\p / \p x^\mu)$. 
The transformation behaviour of the metric tensor, noting that
the Lie derivative for a type (0,2) tensor is given by
\bea
\label{eq:lie_tensor} \pounds_{\xi}g_{\mu\nu}
=g_{\mu\nu,\lambda}\xi^\lambda
+g_{\mu\lambda}\xi^\lambda_{,~\nu}+g_{\lambda\nu}\xi^\lambda_{,~\mu}\,,
\eea
is
\be
\label{eq:metrictransfirst}
\wt{\delta g^{(1)}_{\mu\nu}}=\delta g^{(1)}_{\mu\nu}
+g^{(0)}_{\mu\nu,\lambda}\xi^\lambda_1
+g^{(0)}_{\mu\lambda}\xi^\lambda_{1~,\nu}
+g^{(0)}_{\lambda\nu}\xi^\lambda_{1~,\mu}\,.
\ee
We can obtain the transformation behaviour of each particular
metric function by extracting it, in turn, from the above
general expression using the method outlined in e.g. MW2009.
As mentioned above, we do not focus on details here but instead
quote the results.
The scalar metric perturbations transform as 
\begin{eqnarray}
\widetilde{\iphi}&=\iphi+\H\alpha_1+\alpha_1' \,,\\
\widetilde{\ipsi}&=\ipsi-\H\alpha_1\,, \\
\widetilde{B_1}&=B_1-\alpha_1+\beta_1'\,,\\
\widetilde{E_1}&=E_1+\beta_1\,,
\end{eqnarray}
the vector metric perturbations as
\begin{eqnarray}
\widetilde{S_1{}^i}&=S_1{}^i-\gamma_1{}^{i'}\,,\\
\widetilde{F_1{}^i}&=F_1{}^i+\gamma_1{}^i\,,
\end{eqnarray}
and  the tensor perturbation, $h_{1ij}$, is gauge invariant.
Finally, the scalar shear, which is defined as $\sigma_1=E_1'-B_1$,
transforms as
\be 
\widetilde{\sigma_1}=\sigma_1+\alpha_1\,,
\ee 
which will be useful later when we come to define gauges in
\S\ref{sec:Comparison_of_different_formulations}.

%****************************************************************

%%%%%%%%%%%%%%%%%%%%%%%%%%%%%%%%%%%%%%%%%%%%%%%%%%%%%%%%%%%%%%%%%%%%%%
\subsection{KN2010 formulation}
\label{sec:Nakamura_formulation}
%%%%%%%%%%%%%%%%%%%%%%%%%%%%%%%%%%%%%%%%%%%%%%%%%%%%%%%%%%%%%%%%%%%%%%

%****************************************************************

An alternative approach within perturbation theory was 
presented in KN2010~\cite{Nakamura:2010yg}, where the procedure
proposed by KN in 2003~\cite{kouchan-gauge-inv} is used to
construct gauge invariant variables.
Evaluating Eq.~(\ref{eq:Taylor-exp-of-a-tensor-field-general})
to first-order perturbation, the gauge transformation rule of
the first order metric perturbation $h_{ab}$ is given by
\begin{eqnarray}
  {}_{{\cal Y}}h_{ab} - {}_{{\cal X}}h_{ab}
  =: {\pounds}_{\xi_{1}}g_{ab}.
  \label{eq:first-order-gauge-trans-metric-org}
\end{eqnarray}
We decompose the linear metric perturbation, $h_{ab}$, as
\begin{eqnarray}
  h_{ab} =: {\cal H}_{ab} + {\pounds}_{X}g_{ab},
  \label{eq:linear-metric-decomp}
\end{eqnarray}
where ${\cal H}_{ab}$ and $\pounds_X g_{ab}$ are the gauge
invariant and variant parts of the first order metric
perturbations~\cite{kouchan-gauge-inv}, respectively and $X^a$ is
defined below.
That is, under a gauge transformation, these are transformed as 
\begin{equation}
  {}_{{\cal Y}}\!{\cal H}_{ab} - {}_{{\cal X}}\!{\cal H}_{ab} =  0, 
  \quad
  {}_{\quad{\cal Y}}\!X^{a} - {}_{{\cal X}}\!X^{a} = \xi^{a}_{(1)}. 
  \label{eq:linear-metric-decomp-gauge-trans}
\end{equation}
We note that the decomposition in Eq.~(\ref{eq:linear-metric-decomp})
is an assumption, however for perturbations to a FRW spacetime it can
be shown to be correct~\cite{Nakamura:2010yg}.

%****************************************************************

To proceed, we consider the scalar-vector-tensor decomposition
of the components $h_{i\eta}$, $h_{ij}$ of $h_{ab}$ as 
\begin{eqnarray}
  h_{i\eta} &=& D_{i}h_{(VL)} + h_{(V)i}
  \\
  h_{ij} &=& a^{2} \left\{
    h_{(L)} \delta_{ij}
    + \left(D_{i}D_{j} - \frac{1}{3}\delta_{ij}\Delta\right)h_{(TL)}
    + 2 D_{(i}h_{(TV)j)} + {h_{(TT)ij}}
  \right\}
\end{eqnarray}
where $\Delta:=D^{i}D_{i}=\delta^{ij}D_{i}D_{j}$.
Further $h_{(V)i}$, $h_{(TV)j}$, and ${h_{(TT)ij}}$ satisfy the
properties 
\begin{eqnarray}
  &&
  D^{i}h_{(V)i} = 0, \quad  
  D^{i} h_{(TV)i} = 0, 
  {h_{(T)}}^{i}_{\;\;i} := \delta^{ij}{h_{(T)}}_{ij} = 0,
  \label{eq:bare-parturbative-property}
  \\
  &&
  h_{(TT)ij} = h_{(TT)ji}, \quad
  D^{i} h_{(TT)ij} = 0.
  \nonumber
\end{eqnarray}
The generator of the gauge transformation, $\xi^a$, is also decomposed as
\begin{equation}
  \xi_{a} = \xi_{\eta}(d\eta)_{a}
  + \left(D_{i}\xi_{(L)} + \xi_{(T)i}\right) (dx^{i})_{a}, \quad
   D^{i}\xi_{(T)i} = 0.
  \label{eq:xii-decomp}
\end{equation}
Using Eq.~(\ref{eq:first-order-gauge-trans-metric-org}) we can
then obtain gauge transformation rules for the components of $h_{ab}$:
\begin{eqnarray}
  \label{eq:KN2010-linear-gauge-trans-metric-hetaeta}
  {}_{\cal Y}\!h_{\eta\eta} - {}_{\cal X}\!h_{\eta\eta}
  &=& 2\left(\partial_{\eta} - H\right) \xi_{\eta}
  , \\
  \label{eq:KN2010-decomposed-gauge-trans-2}
  {}_{\cal Y}\!h_{(VL)} - {}_{\cal X}\!h_{(VL)}
  &=& \xi_{\eta} + \left(\partial_{\eta} - 2 H\right) \xi_{(L)}, \\
  \label{eq:KN2010-decomposed-gauge-trans-3}
  {}_{\cal Y}\!h_{(V)i} - {}_{\cal X}\!h_{(V)i}
  &=&
  \left( \partial_{\eta} - 2 H \right) \xi_{(T)i}, \\
  \label{eq:KN2010-decomposed-gauge-trans-4}
  a^{2} {}_{\cal Y}\!h_{(L)} - a^{2} {}_{\cal X}\!h_{(L)}
  &=&
  - 2 H \xi_{\eta} + \frac{2}{3}\Delta\xi_{(L)}, \\
  \label{eq:KN2010-decomposed-gauge-trans-5}
  a^{2} {}_{\cal Y}\!h_{(TL)} - a^{2} {}_{\cal X}\!h_{(TL)}
  &=&
  2 \xi_{(L)} , \\
  \label{eq:KN2010-decomposed-gauge-trans-6}
  a^{2} {}_{\cal Y}\!h_{(TV)i} - a^{2} {}_{\cal X}\!h_{(TV)i}
  &=&
  \xi_{(T)i}, \\
  \label{eq:KN2010-decomposed-gauge-trans-7}
  a^{2} {}_{\cal Y}\!h_{(TT)ij} - a^{2} {}_{\cal X}\!h_{(TT)ij}
  &=& 0.
\end{eqnarray}

%****************************************************************

We then inspect these gauge transformation rules and define
gauge invariant variables.
Firstly, Eq.~(\ref{eq:KN2010-decomposed-gauge-trans-7}) shows
that the transverse-traceless part $h_{(TT)ij}$ is itself gauge 
invariant, as expected.
We denote this as
\begin{eqnarray}
  \label{eq:KN2010-tensor-mode-gauge-inv-def}
  \stackrel{(1)}{\chi}_{ij} := h_{(TT)ij}, 
\end{eqnarray}
Secondly, the gauge-transformation rules
(\ref{eq:KN2010-decomposed-gauge-trans-3}) and
(\ref{eq:KN2010-decomposed-gauge-trans-6}) give the transverse
vector-mode $\stackrel{(1)\;\;}{\nu_{i}}$ defined as
\begin{eqnarray}
  a^{2} \stackrel{(1)\;\;}{\nu_{i}} &:=&
  h_{(V)i} - a^{2}\partial_{\eta}h_{(TV)i}.
  \label{eq:KN2010-vector-mode-gauge-inv-def}
\end{eqnarray}
In addition to the vector and tensor modes 
there are two scalar modes in the first order metric
perturbation, $h_{ab}$.
To see this, we first consider the gauge transformation rules
(\ref{eq:KN2010-decomposed-gauge-trans-2}) and
(\ref{eq:KN2010-decomposed-gauge-trans-5}).
From these transformation rules, the variable $X_{\eta}$ defined
by 
\begin{equation}
  X_{\eta} :=
  h_{(VL)} 
  - \frac{1}{2} a^{2} \partial_{\eta} h_{(TL)} 
  \label{eq:KN2010-barXeta-def}
\end{equation}
transforms as 
\begin{eqnarray}
  \label{eq:KN2010-barXtau-gauge-trans}
  {}_{\cal Y}X_{\eta} - {}_{\cal X}X_{\eta} = \xi_{\eta}. 
\end{eqnarray}
Using this definition of $X_{\eta}$, and inspecting the gauge
transformation rule
(\ref{eq:KN2010-linear-gauge-trans-metric-hetaeta}), we can show
that the variable $\stackrel{(1)}{\Phi}$ defined as 
\begin{equation}
  \label{eq:KN2010-scalar-Phi-gauge-inv-def}
  - 2 a^{2} \stackrel{(1)}{\Phi}
  := 
  h_{\eta\eta} - 2 \left( \partial_{\eta} - H \right) X_{\eta}
\end{equation}
is gauge invariant.
Furthermore, from gauge transformation rules
(\ref{eq:KN2010-decomposed-gauge-trans-4}),
(\ref{eq:KN2010-decomposed-gauge-trans-5}), and
(\ref{eq:KN2010-barXtau-gauge-trans}), the variable
$\stackrel{(1)}{\Psi}$ defined by 
\begin{equation}
  \label{eq:KN2010-scalar-Psi-gauge-inv-def}
  - 2 a^{2} \stackrel{(1)}{\Psi} := 
  a^{2} \left( h_{(L)} - \frac{1}{3}\Delta h_{(TL)} \right)
  + 2 H X_{\eta}
\end{equation}
is gauge invariant.
The set of variables $\{\stackrel{(1)}{\Phi}$,
$\stackrel{(1)}{\Psi}$, $\stackrel{(1)\;\;}{\nu_{i}}$,
$\stackrel{(1)}{\chi}_{ij}\}$ is the complete set of gauge
invariant variables.

%****************************************************************

We can now write the original metric perturbation,
$h_{ab}$, in terms of these gauge invariant variables as
\begin{eqnarray}
  \label{eq:htautau-gaugeinv-decomp}
  h_{\eta\eta} 
  &=& 
  - 2 a^{2} \stackrel{(1)}{\Phi}
  + 2 \left( \partial_{\eta} - H \right) X_{\eta}
  , \\
  h_{\eta i}
  &=& 
    a^{2} \stackrel{(1)\;\;}{\nu_{i}} 
  + a^{2} \partial_{\eta} h_{(TV)i}
  + D_{i} h_{(VL)}
  \label{eq:htaui-gaugeinv-decomp}
  ,\\
  h_{ij} 
  &=&
  - 2 a^{2} \stackrel{(1)}{\Psi} \delta_{ij}
  + a^{2} \stackrel{(1)\;\;}{\chi_{ij}}
  + a^{2} D_{i}D_{j} h_{(TL)}
  - 2 H \bar{X}_{\eta} \delta_{ij}
  + 2 a^{2} D_{(i}h_{(TV)j)}
  .
  \label{eq:hij-gaugeinv-decomp}
\end{eqnarray}
From the gauge invariance of the variables
$\stackrel{(1)}{\chi}_{ij}$, $\stackrel{(1)\;\;}{\nu_{i}}$,
$\stackrel{(1)}{\Phi}$, and $\stackrel{(1)}{\Psi}$, we may
read off the gauge invariant part of $h_{ab}$ as    
\begin{eqnarray}
  {\cal H}_{ab}
  &=&
  a^{2} \left\{
    - 2 \stackrel{(1)}{\Phi} (d\eta)_{a}(d\eta)_{b}
    + 2 \stackrel{(1)}{\nu}_{i} (d\eta)_{(a}(dx^{i})_{b)}
  \right.
  \nonumber\\
  && \quad\quad\quad
  \left.
    + 
    \left( - 2 \stackrel{(1)}{\Psi} \delta_{ij} 
      + \stackrel{(1)}{\chi}_{ij} \right)
    (dx^{i})_{a}(dx^{j})_{b}
  \right\},
  \label{eq:KN2010-components-calHab}
\end{eqnarray}
The remaining gauge dependent parts in
Eqs.~(\ref{eq:htautau-gaugeinv-decomp})--(\ref{eq:hij-gaugeinv-decomp})
should then be given in the form ${\pounds}_{X}g_{ab}$ for some
vector field $X_{a}$.
In fact, such a vector field is
\begin{eqnarray}
  \label{eq:KN2010-X_a-def}
  X_{a}
  :=
  X_{\eta} (d\eta)_{a}
  +
  a^{2} \left(
      h_{(TV)i}
    + \frac{1}{2} D_{i}h_{(TL)}
  \right)
  (dx^{i})_{a},
\end{eqnarray}
where $X_{\eta}$ is defined in
Eq.~(\ref{eq:KN2010-barXeta-def}). 
One can also check that this vector field $X_{a}$ satisfies
Eq.~(\ref{eq:linear-metric-decomp-gauge-trans}).

%*******************************************************************

Thus, it has been shown that the decomposition
(\ref{eq:linear-metric-decomp}) for the first order metric
perturbation is correct for cosmological perturbations.
We should note that, in order to accomplish
Eq.~(\ref{eq:linear-metric-decomp}), we have assumed the
existence of the Green's function $\Delta^{-1}$,
and that the perturbative modes which
belong to the kernel of the operator $\Delta$ have been
neglected. 
However, we ignore both of these issues in this paper, pointing
the diligent reader to, for example,
Refs.~\cite{kouchan-cosmo-second,Urakawa:2010it} for more
information.

%*******************************************************************

Finally, the correspondence 
between the variables for linear perturbations for
the KN2010 approach presented in this section and those
for the MW2009 approach presented in
\S~\ref{sec:standard_formulation} are 
\begin{eqnarray}
  &&
  h_{\eta\eta} \Leftrightarrow - 2 a^{2} \phi_{1}, \quad
  h_{(VL)} \Leftrightarrow a^{2}B_{1}, \quad
  h_{(V)i} \Leftrightarrow - a^{2}S_{1i}
  , \nonumber\\
  &&
  h_{(L)} \Leftrightarrow - 2 \psi_{1} + \frac{2}{3}\Delta E_{1}, \quad
  h_{(TL)} \Leftrightarrow 2 E_{1}, \quad
  h_{(TV)i} \Leftrightarrow F_{1i}, \quad
  h_{(TT)ij} \Leftrightarrow h_{1ij}
  \label{eq:KN2010-MW2009-variables-correspondence}
  , \\
  &&
  \xi_{\eta} \Leftrightarrow - a^{2}\alpha_{1}, \quad
  \xi_{(L)} \Leftrightarrow a^{2} \beta_{1}, \quad
  \xi_{(T)i} \Leftrightarrow a^{2} \gamma_{1i}.
  \nonumber
\end{eqnarray}
Note that at first order there is no 
difference between the two formalisms, as can be seen from
Eq.~(\ref{eq:KN2010-MW2009-variables-correspondence}).

%****************************************************************

%%%%%%%%%%%%%%%%%%%%%%%%%%%%%%%%%%%%%%%%%%%%%%%%%%%%%%%%%%%%%%%%%%%%%%
%%%%%%%%%%%%%%%%%%%%%%%%%%%%%%%%%%%%%%%%%%%%%%%%%%%%%%%%%%%%%%%%%%%%%%
\section{Second order cosmological perturbations}
\label{sec:Formulations_for_the_second-order_cosmological_perturbations}
%%%%%%%%%%%%%%%%%%%%%%%%%%%%%%%%%%%%%%%%%%%%%%%%%%%%%%%%%%%%%%%%%%%%%%
%%%%%%%%%%%%%%%%%%%%%%%%%%%%%%%%%%%%%%%%%%%%%%%%%%%%%%%%%%%%%%%%%%%%%%

%****************************************************************

Having reviewed linear cosmological perturbation theory in both
the MW2009 approach and that in KN2010 above, in this section,
we review the formalisms at second order.

%****************************************************************

%%%%%%%%%%%%%%%%%%%%%%%%%%%%%%%%%%%%%%%%%%%%%%%%%%%%%%%%%%%%%%%%%%%%%%
\subsection{MW2009 formulation}
\label{sec:second_standard}
%%%%%%%%%%%%%%%%%%%%%%%%%%%%%%%%%%%%%%%%%%%%%%%%%%%%%%%%%%%%%%%%%%%%%%

%****************************************************************

In order to consider cosmological perturbation theory to second
order we do not truncate the perturbative expansion after the
first term in Eq.~(\ref{app1}).
The metric tensor then has components
form
\begin{eqnarray}
\label{eq:lineelement-second}
\delta g_{00}^{(2)}&=-2a^2(\eta)\phi_2\,,\\
\delta g_{0i}^{(2)}&=a^2(\eta)(B_{2,i}-S_{2i})\,,\\
\delta g_{ij}^{(2)}&=a^2(\eta)\Big(-2\psi_2\delta_{ij}+2E_{2,ij}+2F_{2(i,j)}+h_{2ij}\Big)\,,
\end{eqnarray}
where the quantities here are analogous to their first order
counterparts.
At second order we split the generating vector $\xi_2^\mu$, as
at first order, as
\be 
\xi_2^\mu=(\alpha_2,\beta_{2,}{}^i+\gamma_2{}^i)\,.
\ee
Then, using \eq{transform}, we find that the second order energy
density perturbation transforms as
\be
\label{eq:rhotrans2}
\widetilde{\delta\rho_2}=\delta\rho_2
+\rhob'\alpha_2+\alpha_1(\rhob''\alpha_1+\rhob'\alpha_1'+2\delta\rho_1')
+(2\delta\rho_1+\rhob'\alpha_1)_{,k}(\beta_{1,}{}^k+\gamma_{1}{}^k)\,,
\ee
where we note for the first time here that, while at linear
order different types of perturbation (scalar, vector and
tensor) decouple, this is no longer true at higher order.
This is a crucial qualitative difference between first and
second order perturbation theory and can lead to the generation
of, for example, second order gravitational waves~\cite{tensors}
or vector modes and vorticity~\cite{vectors}.
At second order the metric tensor transforms as
\begin{eqnarray}
\label{eq:metrictranssecond}
\wt{\delta g^{(2)}_{\mu\nu}}&=\delta g^{(2)}_{\mu\nu}
+g^{(0)}_{\mu\nu,\lambda}\xi^\lambda_2
+g^{(0)}_{\mu\lambda}\xi^\lambda_{2~,\nu}
+g^{(0)}_{\lambda\nu}\xi^\lambda_{2~,\mu}
+2\Big[
\delta g^{(1)}_{\mu\nu,\lambda}\xi^\lambda_1
+\delta g^{(1)}_{\mu\lambda}\xi^\lambda_{1~,\nu}
+\delta g^{(1)}_{\lambda\nu}\xi^\lambda_{1~,\mu}
\Big]\nonumber \\
&\quad+g^{(0)}_{\mu\nu,\lambda\alpha}\xi^\lambda_1\xi^\alpha_1
+g^{(0)}_{\mu\nu,\lambda}\xi^\lambda_{1~,\alpha}\xi^\alpha_1
+2\Big[
g^{(0)}_{\mu\lambda,\alpha} \xi^\alpha_1\xi^\lambda_{1~,\nu}
+g^{(0)}_{\lambda\nu,\alpha} \xi^\alpha_1\xi^\lambda_{1~,\mu}
+g^{(0)}_{\lambda\alpha}  \xi^\lambda_{1~,\mu} \xi^\alpha_{1~,\nu}
\Big]
\nonumber \\
&\quad+g^{(0)}_{\mu\lambda}\left(
\xi^\lambda_{1~,\nu\alpha}\xi^\alpha_1
+\xi^\lambda_{1~,\alpha}\xi^\alpha_{1,~\nu}
\right)
+g^{(0)}_{\lambda\nu}\left(
\xi^\lambda_{1~,\mu\alpha}\xi^\alpha_1
+\xi^\lambda_{1~,\alpha}\xi^\alpha_{1,~\mu}
\right)\,.
\end{eqnarray}
From this we can extract, as at first order, the transformation
behaviour of the metric perturbations.
Again, we refer to MW2009~\cite{Malik:2008im} for the details and
only quote the results in this section. One finds that the
scalar metric functions transform as
\begin{eqnarray}
\label{transphi2}
\widetilde {\iiphi} &= \iiphi+\H\alpha_2+{\alpha_2}'
+\alpha_1\left[{\alpha_1}''+5\H{\alpha_1}' +\left(\H'+2\H^2
\right)\alpha_1 +4\H\phi_1+2\phi_1'\right] \\
&\quad+2{\alpha_1}'\left({\alpha_1}'+2\phi_1\right)
+\xi_{1k}
\left({\alpha_1}'+\H{\alpha_1}+2\phi_1\right)_{,}^{~k}
+\xi_{1k}'\left[\alpha_{1,}^{~k}-2B_{1k}-{\xi_1^k}'\right]\,, \nn \\
\label{transpsi2}
\wt\psi_2&=\psi_2-\H\alpha_2-\frac{1}{4}\X^k_{~k}
+\frac{1}{4}\nabla^{-2} \X^{ij}_{~~,ij}\,,\\
\label{transE2}
\wt E_2&=E_2+\beta_2+\frac{3}{4}\nabla^{-2}\nabla^{-2}\X^{ij}_{~~,ij}
-\frac{1}{4}\nabla^{-2}\X^k_{~k}\,,\\
\label{transB2}
\widetilde B_{2} &= B_{2}-\alpha_2+\beta_2' +\nabla^{-2} \XB{}^k_{~,k}
\,,
\end{eqnarray}
where $\X_{ij}$ and $\XB_i$ are defined as
\begin{eqnarray}
\label{defXBi}
\XB_{i}
&\equiv
2\Big[
\left(2\H B_{1i}+B_{1i}'\right)\alpha_1
+B_{1i,k}\xi_1^k-2\phi_1\alpha_{1,i}+B_{1k}\xi_{1,~i}^k
+B_{1i}\alpha_1'+2 C_{1ik}{\xi_{1}^k}'
 \Big]\nonumber\\
&+4\H\alpha_1\left(\xi_{1i}'-\alpha_{1,i}\right)
+\alpha_1'\left(\xi_{1i}'-3\alpha_{1,i}\right)
+\alpha_1\left(\xi_{1i}''-\alpha_{1,i}'\right)\nonumber\\
&+{\xi_{1}^k}'\left(\xi_{1i,k}+2\xi_{1k,i}\right)
+\xi_{1}^k\left(\xi_{1i,k}'-\alpha_{1,ik}\right)
-\alpha_{1,k}\xi_{1,~i}^k\,,\\
\label{Xijdef}
\X_{ij}&\equiv
2\Big[\left(\H^2+\frac{a''}{a}\right)\alpha_1^2
+\H\left(\alpha_1\alpha_1'+\alpha_{1,k}\xi_{1}^{~k}
\right)\Big] \delta_{ij}+2\left(B_{1i}\alpha_{1,j}+B_{1j}\alpha_{1,i}\right)\nonumber\\
&
+4\Big[\alpha_1\left(C_{1ij}'+2\H C_{1ij}\right)
+C_{1ij,k}\xi_{1}^{~k}+C_{1ik}\xi_{1~~,j}^{~k}
+C_{1kj}\xi_{1~~,i}^{~k}\Big]
\nonumber\\
&
+4\H\alpha_1\left( \xi_{1i,j}+\xi_{1j,i}\right)
-2\alpha_{1,i}\alpha_{1,j}+2\xi_{1k,i}\xi_{1~~,j}^{~k}
+\alpha_1\left( \xi_{1i,j}'+\xi_{1j,i}' \right)
\nonumber\\
&+\xi_{1i,k}\xi_{1~~,j}^{~k}+\xi_{1j,k}\xi_{1~~,i}^{~k}
+\xi_{1i}'\alpha_{1,j}+\xi_{1j}'\alpha_{1,i}
+\left(\xi_{1i,jk}+\xi_{1j,ik}\right)\xi_{1}^{~k}
\,.
\end{eqnarray}
We furthermore find that the second order metric vector
perturbations transform as
\begin{eqnarray}
\label{transS2}
\widetilde S_{2i}&=S_{2i}-\gamma_2{}_i{}'-\XB_i+\nabla^{-2}\XB^k_{~,ki}
\,,\\
\label{transFi2}
\wt F_{2i} &= F_{2i}+\gamma_{2i}
+\nabla^{-2}\X_{ik,}^{~~~k}-\nabla^{-2}\nabla^{-2}\X^{kl}_{~~,kli}
\,,
\end{eqnarray}
and the tensor perturbation as
\bea
\label{transhij2}
\wt h_{2ij}&=& h_{2ij}+\X_{ij}
+\frac{1}{2}\left(\nabla^{-2}\X^{kl}_{~~,kl}-\X^k_{~k}
\right)\delta_{ij}
+\frac{1}{2}\nabla^{-2}\nabla^{-2}\X^{kl}_{~~,klij}\nonumber\\
&&+\frac{1}{2}\nabla^{-2}\X^k_{~k,ij}
-\nabla^{-2}\left(\X_{ik,~~~j}^{~~~k}+\X_{jk,~~~i}^{~~~k}
\right)
\,.
\eea
Note that the second order tensor perturbation, $h_{2ij}$, changes 
under a gauge-transformation ~\cite{Malik:2008yp}, unlike the
first order tensor perturbation, $h_{1ij}$.  However, we show that the
expression (\ref{transhij2}) is itself gauge-invariant as discussed at
the end of the next section. Hence, using Eq.~(\ref{transhij2}), we can
construct gauge-invariant tensor perturbations at second order

%****************************************************************

%%%%%%%%%%%%%%%%%%%%%%%%%%%%%%%%%%%%%%%%%%%%%%%%%%%%%%%%%%%%%%%%%%%%%%
\subsection{KN2010 formulation}
\label{sec:second_Nakamura}
%%%%%%%%%%%%%%%%%%%%%%%%%%%%%%%%%%%%%%%%%%%%%%%%%%%%%%%%%%%%%%%%%%%%%%

%****************************************************************

We now consider the construction of the gauge-invariant
variables for second-order metric perturbation $l_{ab}$ in
KN2010~\cite{Nakamura:2010yg}.
As we have confirmed  the decomposition of the first-order
metric perturbation, $h_{ab}$, into the gauge-invariant 
${\cal H}_{ab}$ and gauge-variant parts $X_{a}$, we can now also
find gauge invariant variables for higher-order
perturbations~\cite{kouchan-gauge-inv}.
Using Eq.~(\ref{transform}), the gauge transformation rule for
$l_{ab}$ is 
\begin{eqnarray}
  {}_{{\cal Y}}l_{ab} - {}_{{\cal X}}l_{ab}
  &=& 
  + 2 \pounds_{\xi_{1}}{}_{{\cal X}}h_{ab}
  + \left(
    {\pounds}_{\xi_{2}} + \pounds_{\xi_{1}}^{2}
  \right) g_{ab}.
\end{eqnarray}
Through the first order gauge-variant variable $X_{a}$, we first
define the tensor field $\hat{L}_{ab}$ as 
\begin{eqnarray}
  {}_{{\cal X}}\hat{L}_{ab}
  :=
  {}_{{\cal X}}l_{ab}
  - 2 \pounds_{{}_{{\cal X}}X}{}_{{\cal X}}h_{ab}
  + \pounds_{{}_{{\cal X}}X}^{2}g_{ab}
\end{eqnarray}
so that it transforms as
${}_{{\cal Y}}\hat{L}_{ab}-{}_{{\cal X}}\hat{L}_{ab}={\pounds}_{\sigma}g_{ab}$,
where $\sigma^{a}:=\xi_{2}^{a}+\left[\xi_{1},{}_{{\cal X}}X\right]^{a}$.
Since $\sigma^{a}$ is an arbitary vector field on ${\cal M}_{0}$, 
this gauge-transformation rule is the same as that for the
first order metric perturbation, $h_{ab}$
(\ref{eq:first-order-gauge-trans-metric-org}).
Using a similar procedure to that with which we decomposed
$h_{ab}$, the variable $\hat{L}_{ab}$ is then
$\hat{L}_{ab}=:{\cal L}_{ab}+{\pounds}_{Y}g_{ab}$.
Thus, the decomposition of the second order metric perturbation,
$l_{ab}$, is 
\begin{eqnarray}
  \label{eq:L-ab-in-gauge-X-def-second-1}
  l_{ab}
  =:
  {\cal L}_{ab} + 2 {\pounds}_{X} h_{ab}
  + \left(
      {\pounds}_{Y}
    - {\pounds}_{X}^{2} 
  \right)
  g_{ab},
\end{eqnarray}
where ${\cal L}_{ab}$ and $Y^{a}$ are the gauge invariant and
variant parts of the second order metric perturbations, i.e.,
\begin{eqnarray}
  {}_{{\cal Y}}\!{\cal L}_{ab} - {}_{{\cal X}}\!{\cal L}_{ab} = 0,
  \quad
  {}_{{\cal Y}}\!Y^{a} - {}_{{\cal X}}\!Y^{a}
  = \xi_{(2)}^{a} + [\xi_{(1)},{}_{{\cal X}}X]^{a}.
  \label{eq:gauge-trans-of-calL-and-Y}
\end{eqnarray}

%****************************************************************

We may then choose the components of the gauge invariant
variables ${\cal L}_{ab}$ in
Eq.~(\ref{eq:L-ab-in-gauge-X-def-second-1}) as 
\begin{eqnarray}
  {\cal L}_{ab}
  &=&
  a^{2} \left\{
    - 2 \stackrel{(2)}{\Phi} (d\eta)_{a}(d\eta)_{b}
    + 2 \stackrel{(2)}{\nu}_{i} (d\eta)_{(a}(dx^{i})_{b)}
  \right.
  \nonumber\\
  && \quad\quad\quad
  \left.
    + 
    \left( - 2 \stackrel{(2)}{\Psi} \delta_{ij} 
      + \stackrel{(2)}{\chi}_{ij} \right)
    (dx^{i})_{a}(dx^{j})_{b}
  \right\},
  \label{eq:second-order-gauge-inv-metrc-pert-components}
\end{eqnarray}
where $\stackrel{(2)}{\nu}_{i}$ and
$\stackrel{(2)\;\;\;\;}{\chi_{ij}}$ satisfy the equations
\begin{eqnarray}
  && D^{i}\stackrel{(2)\;\;}{\nu_{i}} = 0, \quad
  \stackrel{(2)\;\;\;\;}{\chi^{i}_{\;\;i}} = 0, \quad
   D^{i}\stackrel{(2)\;\;\;\;}{\chi_{ij}} = 0.
\end{eqnarray}
The gauge invariant variables $\stackrel{(2)}{\Phi}$ and
$\stackrel{(2)}{\Psi}$ are the second order scalar perturbations 
and $\stackrel{(2)\;\;}{\nu_{i}}$ and
$\stackrel{(2)\;\;\;\;}{\chi_{ij}}$ are the second order vector
and tensor modes of the metric perturbations, respectively.
Furthermore, using $X^{a}$ and $Y^{a}$, the
gauge invariant variables for an arbitrary tensor field $Q$
can then be defined:
\begin{eqnarray}
  \label{eq:matter-gauge-inv-def-1.0}
  {}^{(1)}\!{\cal Q} &:=& {}^{(1)}\!Q - {\pounds}_{X}Q_{0}
  , \\ 
  \label{eq:matter-gauge-inv-def-2.0}
  {}^{(2)}\!{\cal Q} &:=& {}^{(2)}\!Q - 2 {\pounds}_{X} {}^{(1)}Q 
  - \left\{ {\pounds}_{Y} - {\pounds}_{X}^{2} \right\} Q_{0}
  .
\end{eqnarray}

%****************************************************************

Thus, from Eqs.~(\ref{eq:matter-gauge-inv-def-1.0}) and
(\ref{eq:matter-gauge-inv-def-2.0}), we can see that any 
variable can be decomposed into a gauge invariant and 
gauge variant part as
\begin{eqnarray}
  \label{eq:matter-gauge-inv-decomp-first}
  {}^{(1)}\!Q &=& {}^{(1)}\!{\cal Q} + {\pounds}_{X}Q_{0}
  , \\ 
  \label{eq:matter-gauge-inv-decomp-second}
  {}^{(2)}\!Q &=& {}^{(2)}\!{\cal Q} + 2 {\pounds}_{X} {}^{(1)}Q 
  + \left\{ {\pounds}_{Y} - {\pounds}_{X}^{2} \right\} Q_{0}
  .
\end{eqnarray}
These decomposition formulae are valid for the perturbations of
an arbitrary tensor field without knowing detailed information
of the background metric $g_{ab}^{(0)}$. 
As a collorary, any equation for the perturbations (e.g.~the
Einstein equations or the equations of motion for the matter
fields) is automatically given in a gauge invariant
form~~\cite{kouchan-second,kouchan-second-cosmo-matter} due to
the lower order equations.

%****************************************************************

Since we know that any of these equations must be gauge invariant, we
only need study the gauge invariant parts of
Eq.~(\ref{eq:L-ab-in-gauge-X-def-second-1}) and not treat the
components of $l_{ab}$ directly: that is, we must only study 
${\cal L}_{ab}$.
However, the choice of the gauge invariant part at both first
order, ${\cal H}_{ab}$, and second order, ${\cal L}_{ab}$, is
not unique. 
This can be seen as follows: through the components of
${\cal H}_{ab}$, we can construct a gauge invariant vector field,
for example,
$Z_{a}:=-a\stackrel{(1)}{\Phi}(d\eta)_{a}+a\stackrel{(1)}{\nu}_{i}(dx^{i})_{a}$.
Using this we may write 
\begin{eqnarray}
  h_{ab}
  =
  {\cal H}_{ab}
  + {\pounds}_{Z}g_{ab}
  - {\pounds}_{Z}g_{ab}
  + {\pounds}_{X}g_{ab}
  =:
  {\cal K}_{ab} + {\pounds}_{\tilde{X}}g_{ab},
  \label{eq:non-uniqueness-of-gauge-invariant-variables-first}
\end{eqnarray}
where we have defined 
${\cal K}_{ab}:={\cal H}_{ab}+{\pounds}_{Z}g_{ab}$ and 
$\tilde{X}^{a}:=X^{a}-Z^{a}$.
Thus there are, in principle, infinitely many choices of 
${\cal H}_{ab}$, which correspond to the infinitely many choices
of the gauge fixing. 
As mentioned in a previous paper~\cite{Nakamura:2010yg}, the
situation at second order is more complicated; we simply note
here that there exist infinitely many choices of ${\cal L}_{ab}$.
Because of this, it is evident that relating the variables
in the MW2009 approach to those in the KN2010 approach is more
difficult.
However, the procedure is the same as at first order:
compare the components of $l_{ab}$ through
Eq.~(\ref{eq:L-ab-in-gauge-X-def-second-1}) with 
Eq.~(\ref{eq:lineelement-second}).

%****************************************************************

%%%%%%%%%%%%%%%%%%%%%%%%%%%%%%%%%%%%%%%%%%%%%%%%%%%%%%%%%%%%%%%%%%%%%%
%%%%%%%%%%%%%%%%%%%%%%%%%%%%%%%%%%%%%%%%%%%%%%%%%%%%%%%%%%%%%%%%%%%%%%
\section{Comparison of different formulations}
\label{sec:Comparison_of_different_formulations}
%%%%%%%%%%%%%%%%%%%%%%%%%%%%%%%%%%%%%%%%%%%%%%%%%%%%%%%%%%%%%%%%%%%%%%
%%%%%%%%%%%%%%%%%%%%%%%%%%%%%%%%%%%%%%%%%%%%%%%%%%%%%%%%%%%%%%%%%%%%%%

%****************************************************************

Having now summarised cosmological perturbations from both the
MW2009 approach, and that in KN2010, we move to the main
point of this paper: linking the two approaches and showing the
equivalence between them.

%****************************************************************

%%%%%%%%%%%%%%%%%%%%%%%%%%%%%%%%%%%%%%%%%%%%%%%%%%%%%%%%%%%%%%%%%%%%%%
\subsection{First order}
%%%%%%%%%%%%%%%%%%%%%%%%%%%%%%%%%%%%%%%%%%%%%%%%%%%%%%%%%%%%%%%%%%%%%%

%****************************************************************

In this section we present the comparison between the two
approaches at linear order.
As shown in Eqs.~(\ref{eq:KN2010-MW2009-variables-correspondence}), at
linear order, the correspondence between metric perturbation variables 
in the MW2009 and KN2010 approach is quite clear.
Here we discuss the correspondence of the approaches themselves
by choosing two popular gauges: the longitudinal (or Poisson)
gauge, and the uniform curvature gauge.

%****************************************************************

%%%%%%%%%%%%%%%%%%%%%%%%%%%%%%%%%%%%%%%%%%%%%%%%%%%%%%%%%%%%%%%%%%%%%%
\subsubsection{Longitudinal (Poisson) gauge}

%****************************************************************

The longitudinal gauge is defined in perturbation theory by
choosing hypersurfaces of constant shear 
($\wt \sigma_{1\lg}=0$).
Thus, in the longitudinal gauge, the scalar gauge function
$\alpha_{1\lg}$ is given by
\be 
\label{eq:alpha1long}
\alpha_{1 \lg }=-\sigma_1=B_1-E_1' \,,
\ee 
and this gauge is fully specified (for scalars) by requiring
separately that $\wt{E_{1\lg}}=0$ (which implies that
$\wt{B_{1\lg}}=0$) and is a natural choice.
Hence 
\be 
\label{eq:beta1long}
\beta_{1\lg}=-E_1\,.
\ee
The remaining two scalar metric perturbations, $\iphi$ and
$\ipsi$, are then given as
\begin{eqnarray}
\label{defphi1l}
\wt{\phi_{1\lg}} &= \iphi + \H(B_1-E_1') + (B_1-E_1')' \,, \\
\label{defpsi1l} 
\wt{\psi_{1\lg}} &= \ipsi - \H \left( B_1-E_1'\right) \,.
\end{eqnarray}
These are the two gauge invariant Bardeen potentials $\Phi$ and
$\Psi$~\cite{Bardeen-1980} and, in fact, these two variables
coincide with $\stackrel{(1)}{\Phi}$ and $\stackrel{(1)}{\Psi}$
as defined in KN2010~\cite{Nakamura:2010yg}.

%****************************************************************

By including vector perturbations (an extension of the
longitudinal gauge generally called the Poisson gauge) we must
also fix the vector gauge function $\gamma_1{}^i$, which can be
achieved through the relationship
\be 
\label{eq:gamma1long}
\gamma_{1\lg}^i=\int S_1^id\eta+{\mathcal{C}}_1^i(x^j)\,,
\ee 
where ${\mathcal{C}}_1^i(x^j)$ is an arbitrary constant 3-vector
which depends upon the choice of spatial coordinates on an
initial hypersurface.

%****************************************************************

We have thus specified the gauge generating vector, $\xi_1^\mu$
through Eqs. (\ref{eq:alpha1long}), (\ref{eq:beta1long}), and
(\ref{eq:gamma1long}).
In this gauge, the remaining components of the
linear order metric perturbation are completely given in
the form Eq.~(\ref{eq:KN2010-components-calHab}).
Therefore, this gauge-fixing corresponds to the choice of the
gauge variant part $X^{a}$ in
Eq.~(\ref{eq:linear-metric-decomp}) so that
\begin{eqnarray}
  \label{eq:first-order-longitudinal-gauge-choice-X}
  X^{a}={}_{{\cal P}}X^{a}=0 ,
\end{eqnarray}
where ${\cal P}$ denotes the Poisson (longitudinal) gauge
choice. 
In other words, the gauge invariant variables used in the KN2010
approach~\cite{Nakamura:2010yg} correspond to the variables
associated with the longitudinal gauge.

%****************************************************************

%%%%%%%%%%%%%%%%%%%%%%%%%%%%%%%%%%%%%%%%%%%%%%%%%%%%%%%%%%%%%%%%%%%%%%
\subsubsection{Uniform curvature (spatially flat) gauge}

%****************************************************************

An alternative gauge choice is the uniform curvature, or
spatially flat gauge.
This amounts to choosing a spatial hypersurface on which the
metric is unperturbed by scalar or vector perturbations, which
requires $\widetilde{\ipsi}_\fg=\widetilde{E_1}_\fg=0$ and
$\widetilde{F_{1}^i}_\fg={\mathbf 0}$. 
This gives the gauge transformation
\be 
\alpha_{1\fg}=\frac{\ipsi}{\H}\,,
\hspace{5mm}
\beta_{1\fg}=-E_1\,,
\hspace{5mm}
\gamma_{1\fg}^i=-F_1^i\,.
\ee 
The remaining scalar metric perturbations are then 
\begin{eqnarray}
\wt{\phi{}_{1\fg}} &=& \phi_1 + \psi_1 + \left( \frac{\psi_1}{\H}
\right)^{\prime} \,, \\
\wt{B_{1\fg}} &=& B_1-E_1'-\frac{\psi_1}{\H}  \, ,
\end{eqnarray}
which are gauge invariant.
Thus, in the uniform curvature gauge, the metric perturbations
as expressed in Eqs.~(\ref{eq:hetaeta}), (\ref{eq:hieta}) and
(\ref{eq:hij}) are 
\begin{eqnarray}
\label{eq:hetaetaflat}
\wt{h_{\fg}}{}_{\eta\eta} &=&-2a^2\wt{\phi_{1\fg}}\,,\\
\label{hietaflat}
\wt{h_{\fg}}{}_{i\eta} &=& 2a^2\Big(\wt{B_{1\fg}}{}_{,i}-\wt{S_{1\fg i}}\Big)\,,\\
\label{eq:hijflat}
\wt{h_{\fg}}{}_{ij}&=&a^2h_{1ij}\,.
\end{eqnarray}
Now, we compare this with the KN2010 formalism, in which the
gauge choice is regarded as a choice of the gauge variant part
$X_{a}={}_{{\cal F}}X_{a}$, where ${\cal F}$ denotes the flat
gauge choice.
From Eqs.~(\ref{eq:linear-metric-decomp}),
(\ref{eq:KN2010-components-calHab}), and
Eqs.~(\ref{eq:hetaetaflat})--(\ref{eq:hijflat}) we obtain 
\begin{eqnarray}
  &&
  - 2 a^{2} \stackrel{(1)}{\Phi}
  + 2 \partial_{\eta}({}_{{\cal F}}\!X_{\eta})
  - 2 H ({}_{{\cal F}}\!X_{\eta})
  =
  - 2 a^{2}\wt{\phi_{1\fg}}
  \label{eq:MW2009-KN2010-compare-flat-hetaeta}
  \,, \\
  &&
    a^{2} \stackrel{(1)}{\nu}_{i}
  + D_{i}({}_{{\cal F}}\!X_{\eta})
  + \partial_{\eta}({}_{{\cal F}}\!X_{i})
  - 2 H ({}_{{\cal F}}\!X_{i})
  = 2a^2\Big(\wt{B_{1\fg}}{}_{,i}-\wt{S_{1\fg i}}\Big)
  \label{eq:MW2009-KN2010-compare-flat-hieta}
  , \\
  &&
  - 2 a^{2} \stackrel{(1)}{\Psi} \delta_{ij} 
  + a^{2} \stackrel{(1)}{\chi}_{ij}
  + D_{i}({}_{{\cal F}}\!X_{j})
  + D_{j}({}_{{\cal F}}\!X_{i})
  - 2 H \delta_{ij} ({}_{{\cal F}}\!X_{\eta})
  =
  a^{2} h_{1ij}
  \label{eq:MW2009-KN2010-compare-flat-hij}
\end{eqnarray}
Since the transverse traceless part of both sides of 
Eq.~(\ref{eq:MW2009-KN2010-compare-flat-hij}) should coincide
with one another, we may identify 
\begin{eqnarray}
  \label{eq:MW2009-KN2010-correspondence-chiij-h1ij-flat}
  \stackrel{(1)}{\chi}_{ij} = h_{1ij}.
\end{eqnarray}
Further, the trace and the longitudinal part of
Eq.~(\ref{eq:MW2009-KN2010-compare-flat-hij}) is given by 
\begin{eqnarray}
  \label{eq:MW2009-KN2010-compare-flat-hij-trace}
  &&
  - a^{2} \stackrel{(1)}{\Psi}
  + \frac{1}{3}  D^{i}({}_{{\cal F}}\!X_{i})
  - H ({}_{{\cal F}}\!X_{\eta})
  =
  0
  , \\
  \label{eq:MW2009-KN2010-compare-flat-hij-long}
  &&
    D_{i}({}_{{\cal F}}\!X_{j})
  + D_{j}({}_{{\cal F}}\!X_{i})
  =
  0
  \,,
\end{eqnarray}
from which we can choose 
\begin{eqnarray}
  \label{eq:calFXi-choice}
  {}_{{\cal F}}\!X_{i} = 0.
\end{eqnarray}
Further, Eq.~(\ref{eq:MW2009-KN2010-compare-flat-hij-trace})
yields 
\begin{eqnarray}
  \label{eq:calFXeta-choice}
  {}_{{\cal F}}\!X_{\eta}
  =
  - \frac{a^{2}}{H} \stackrel{(1)}{\Psi}.
\end{eqnarray}
Then, identifying the divergenceless part of the both sides of
Eq.~(\ref{eq:MW2009-KN2010-compare-flat-hieta}) with one another
gives 
\begin{eqnarray}
  \label{eq:MW2009-KN2010-correspondence-S1i-nu1i-flat}
  \wt{S_{1\fg\, i}}
  &=&
  - \stackrel{(1)}{\nu}_{i}\,.
\end{eqnarray}
Finally, the scalar parts of
Eq.~(\ref{eq:MW2009-KN2010-compare-flat-hieta}) and
Eq.~(\ref{eq:MW2009-KN2010-compare-flat-hetaeta}) yield
\begin{eqnarray}
  &&
  \wt{B_{1\fg}}
  =
  \frac{1}{a^{2}} {}_{{\cal F}}\!X_{\eta}
  =
  - \frac{1}{H} \stackrel{(1)}{\Psi}
  \label{eq:MW2009-KN2010-correspondence-B1-Psi-flat}
  \,, \\
  &&
  \wt{\phi_{1\fg}}
  =
  \stackrel{(1)}{\Phi}
  + \left(
    1 - \frac{\partial_{\eta}H}{H^{2}}
  \right) \stackrel{(1)}{\Psi}
  + \frac{1}{H} \partial_{\eta}\stackrel{(1)}{\Psi}
  \,.
  \label{eq:MW2009-KN2010-correspondence-phi1flat-Phi}
\end{eqnarray}
As expected, this is simply the relationship between the scalar
metric perturbations in the  uniform curvature gauge and the
Bardeen potentials, the scalar metric perturbations in the
longitudinal gauge.

%****************************************************************

%%%%%%%%%%%%%%%%%%%%%%%%%%%%%%%%%%%%%%%%%%%%%%%%%%%%%%%%%%%%%%%%%%%%%%
\subsection{Second order}
%%%%%%%%%%%%%%%%%%%%%%%%%%%%%%%%%%%%%%%%%%%%%%%%%%%%%%%%%%%%%%%%%%%%%%

%****************************************************************

%%%%%%%%%%%%%%%%%%%%%%%%%%%%%%%%%%%%%%%%%%%%%%%%%%%%%%%%%%%%%%%%%%%%%%
\subsubsection{Longitudinal (Poisson) gauge}

%****************************************************************

In order to extend the Poisson gauge to second order, we
continue as at first order.
First, we require that $\wt{E_2}_\lg=0$, which fixes the scalar
part of the spatial gauge as 
\be
\beta_{2\lg}=- E_2 - \frac{3}{4}\nabla^{-2}\nabla^{-2}\X^{ij}_{~~,ij}
+\frac{1}{4}\nabla^{-2}\X^k_{~k}\,.
\ee
Then, requiring that $\wt{B_2}_\lg=0$ sets $\alpha_{2\lg}$
through Eq.~(\ref{transB2}), and requiring that the vector
$\wt{F_{2}^i}_\lg={\mathbf 0}$ can be used to fix the vector
part of the spatial gauge transformation, up to a constant of
integration as at linear order.
The second order analogues of the gauge invariant Bardeen
potentials $\Phi$ and $\Psi$ are then given as 
\bea
\label{transphi2l}
\widetilde {\phi_{2\lg}} &=& \phi_2+\H\alpha_{2\lg}+{\alpha_{2\lg}}'
+\alpha_{1\lg}\left[{\alpha_{1\lg}}''+5\H{\alpha_{1\lg}}' +\left(\H'+2\H^2
\right)\alpha_{1\lg} +4\H\phi_1+2\phi_1'\right]\nonumber \\
&&+2{\alpha_{1\lg}}'\left({\alpha_{1\lg}}'+2\phi_1\right)
+\xi_{1\lg k}
\left({\alpha_{1\lg}}'+\H{\alpha_{1\lg}}+2\phi_1\right)_{,}^{~k}
+\xi_{1\lg k}'\left[\alpha_{1\lg,}^{~k}-2B_{1k}-{\xi_{1\lg}^k}'\right]\,.\nn\\
\label{transpsi2l}
\wt{\psi_{2\lg}} &=& \psi_2-\H\alpha_{2\lg}-\frac{1}{4}\X^{~k}_{\lg k}
+\frac{1}{4}\nabla^{-2} \X^{~ij}_{\lg~~,ij}\,,
\eea
where $\X_{\lg ij}$ is denotes the quadratic first order terms
in Eq.~(\ref{Xijdef}) using the longitudinal gauge
transformation components $\alpha_{1\lg}$ and $\xi_{1 \lg}^i$.

%****************************************************************

To summarise, the second order metric perturbation has the
components
\begin{eqnarray}
  \label{eq:longitudinal-final-2nd-metric-eta-eta}
  \tilde{l}_{(lg)\eta\eta} &=& - 2 a^{2} \widetilde {\phi_{2\lg}}
  , \\
  \label{eq:longitudinal-final-2nd-metric-i-eta}
  \tilde{l}_{(lg)i\eta} &=& - a^{2} \widetilde{S_{2i}}
  , \\
  \label{eq:longitudinal-final-2nd-metric-i-j}
  \tilde{l}_{(lg)\eta\eta}
  &=&
  a^2(\eta)\Big(-2\widetilde{\psi_2}\delta_{ij}+\wt h_{2ij}\Big)\,.
\end{eqnarray}
As at first order, the choice of the Poisson gauge in the
MW2009 approach corresponds to the choice
\begin{eqnarray}
  \label{eq:first-order-longitudinal-gauge-choice-Y}
  Y^{a}={}_{{\cal P}}Y^{a}=0\,.
\end{eqnarray}
in the KN2010 approach.
Thus, we again confirmed that the gauge invariant variables used
in the KN2010 approach~\cite{Nakamura:2010yg} correspond to the
variables associated with the longitudinal gauge, with the
correspondence between the variables of the MW2009
formulation and KN2010 formulation as
\begin{eqnarray}
  \stackrel{(2)}{\Phi} \Leftrightarrow \widetilde{\phi_{2\lg}}, \quad
  \stackrel{(2)}{\nu}_{i} \Leftrightarrow \widetilde{S_{2i}}, \quad
  \stackrel{(2)}{\Psi} \Leftrightarrow \widetilde{\psi_2}, \quad
  \stackrel{(2)}{\chi}_{ij} \Leftrightarrow \widetilde{h_{2ij}}
\end{eqnarray}

%****************************************************************

%%%%%%%%%%%%%%%%%%%%%%%%%%%%%%%%%%%%%%%%%%%%%%%%%%%%%%%%%%%%%%%%%%%%%%
\subsubsection{Uniform curvature (spatially flat) gauge}

%****************************************************************

At second order, the gauge condition $\wt{\psi_2}=0$ gives,
using 
\eq{transpsi2},
\bea
 \label{deffg2}
\alpha_{2\fg}=\frac{\psi_2}{\H}+\frac{1}{4\H}\left[
\nabla^{-2}\X^{ij}_{\fg,ij}-\X^k_{\fg k}\right]\,, \eea
where we get $\X_{\fg ij}$ from \eq{Xijdef} using the first
order gauge generators given above, as
\bea
\label{Xijflat}
\X_{\fg ij}&=&2\left[
\psi_1\left(\frac{\psi_1'}{\H}+2\psi_1\right)+\psi_{1,k}\xi_{1\fg}^k
\right]\delta_{ij}
+\frac{4}{\H}\psi_1\left(C_{1ij}'+2\H C_{1ij}\right)\nonumber\\
&& +4 C_{1ij,k}\xi_{1\fg}^k + \left(4 C_{1ik}+\xi_{1\fg
i,k}\right)\xi_{1\fg,j}^k
+ \left(4 C_{1jk}+\xi_{1\fg j,k}\right)\xi_{1\fg,i}^k\nonumber\\
&&
+\frac{1}{\H}\Big[
\psi_{1,i}\left(2B_{1j}+\xi_{1\fg j}'\right)
+\psi_{1,j}\left(2B_{1i}+\xi_{1\fg i}'\right)
\Big]
-\frac{2}{\H^2}\psi_{1,i}\psi_{1,j}\nonumber\\
&&
+\frac{2}{\H}\psi_1\left(
\xi_{1\fg (i,j)}'+4\H  \xi_{1\fg (i,j)}\right)
+2\xi_{1\fg}^k\xi_{1\fg (i,j)k}
+2\xi_{1\fg k,i}\xi_{1\fg,j}^k\,,\nonumber\\
\eea
where we define
\be
\xi_{1\fg i}=-\left(E_{1,i}+F_{1i}\right)\,.
\ee 
Finally, the gauge conditions $\wt{E_{2\fg}}=0$ and
$\widetilde{F_{2}^i}_\fg={\mathbf 0}$ enable us to specify the
gauge functions $\beta_{2\fg}$ and $\gamma_{2\fg}^i$
completely. 
Thus, in the uniform curvature gauge, the metric perturbation
has the components 
\begin{eqnarray}
  \label{eq:final-flat-gauge-second-metric-eta-eta}
  \tilde{l}_{(\fg)\eta\eta} &=& - 2 a^{2} \wt{\phi_{2\fg}},\\
  \label{eq:final-flat-gauge-second-metric-i-eta}
  \tilde{l}_{(\fg)\eta i} &=&
  a^2(\eta)\left(
    \widetilde{B_{2\fg}}_{,i} - \widetilde{S_{2\fg i}}
  \right)\,,\\
  \label{eq:final-flat-gauge-second-metric-i-j}
  \tilde{l}_{(\fg)ij} &=&
  a^2(\eta) \widetilde{h_{2\fg ij}}\,,
\end{eqnarray}
On the other hand, according to the decomposition
(\ref{eq:L-ab-in-gauge-X-def-second-1}), the second-order
metric perturbation $l_{ab}$ in the flat gauge is given by 
\begin{eqnarray}
  \label{eq:L-ab-in-gauge-X-def-second-flat-gauge}
  {}_{{\cal F}}\!l_{ab}
  =:
  {\cal L}_{ab} + 2 {\pounds}_{{}_{{\cal F}}\!X} {}_{{\cal F}}\!h_{ab}
  + \left(
      {\pounds}_{{}_{{\cal F}}\!Y}
    - {\pounds}_{{}_{{\cal F}}\!X}^{2} 
  \right)
  g_{ab}.
\end{eqnarray}
Here, the components of ${}_{{\cal F}}\!X_{a}$ are given by 
Eqs.~(\ref{eq:calFXi-choice}) and (\ref{eq:calFXeta-choice}), 
and the components of ${}_{{\cal F}}\!h_{ab}$ are given by
Eqs.~(\ref{eq:hetaetaflat})--(\ref{eq:hijflat}) with the
relations
(\ref{eq:MW2009-KN2010-correspondence-chiij-h1ij-flat}),
(\ref{eq:MW2009-KN2010-correspondence-S1i-nu1i-flat}),
(\ref{eq:MW2009-KN2010-correspondence-B1-Psi-flat}), and
(\ref{eq:MW2009-KN2010-correspondence-phi1flat-Phi}). 
Tedious calculations show that the components of the terms
\begin{eqnarray}
  {}_{{\cal F}}\!\Xi_{ab} 
  :=
  2 {\pounds}_{{}_{{\cal F}}\!X} {}_{{\cal F}}\!h_{ab}
  - {\pounds}_{{}_{{\cal F}}\!X}^{2}g_{ab} 
\end{eqnarray}
in Eq.~(\ref{eq:L-ab-in-gauge-X-def-second-flat-gauge}) are
\begin{eqnarray}
  {}_{{\cal F}}\!\Xi_{\eta\eta}
  &:=&
  \frac{2a^{2}}{H^{4}} \left(
    - 4 H^{4} \stackrel{(1)}{\Phi} \stackrel{(1)}{\Psi}
    - 2 H^{4} (\stackrel{(1)}{\Psi})^{2}
    - 2 H^{2} (\partial_{\eta}\stackrel{(1)}{\Psi})^{2}
    - 2 H^{3} \stackrel{(1)}{\Psi} \partial_{\eta}\stackrel{(1)}{\Phi}
  \right.
  \nonumber\\
  && \quad\quad\quad
  \left.
    - 4 H^{3} \stackrel{(1)}{\Phi} \partial_{\eta}\stackrel{(1)}{\Psi}
    - 5 H^{3} \stackrel{(1)}{\Psi} \partial_{\eta}\stackrel{(1)}{\Psi}
    +   H \partial_{\eta}^{2}H (\stackrel{(1)}{\Psi})^{2}
    - 4 (\partial_{\eta}H)^{2} (\stackrel{(1)}{\Psi})^{2}
  \right.
  \nonumber\\
  && \quad\quad\quad
  \left.
    + 4 H^{2} \partial_{\eta}H \stackrel{(1)}{\Phi} \stackrel{(1)}{\Psi}
    + 4 H^{2} \partial_{\eta}H (\stackrel{(1)}{\Psi})^{2}
    + 6 H \partial_{\eta}H \stackrel{(1)}{\Psi} \partial_{\eta}\stackrel{(1)}{\Psi}
  \right.
  \nonumber\\
  && \quad\quad\quad
  \left.
    -   H^{2} \stackrel{(1)}{\Psi} \partial_{\eta}^{2}\stackrel{(1)}{\Psi}
  \right)
  ,
  \label{eq:calFXi-eta-eta-def}
\end{eqnarray}
\begin{eqnarray}
  {}_{{\cal F}}\!\Xi_{i\eta}
  &:=&
  \frac{a^{2}}{H^{3}} \left[
       8 \partial_{\eta}H \stackrel{(1)}{\Psi} D_{i}\stackrel{(1)}{\Psi}
    -  3 H \stackrel{(1)}{\Psi} \partial_{\eta}D_{i}\stackrel{(1)}{\Psi}
    -  5 H \partial_{\eta}\stackrel{(1)}{\Psi} D_{i}\stackrel{(1)}{\Psi}
  \right.
  \nonumber\\
  && \quad\quad\quad\quad\quad
  \left.
    -  4 H^{2} \stackrel{(1)}{\Phi} D_{i}\stackrel{(1)}{\Psi}
    -  8 H^{2} \stackrel{(1)}{\Psi} D_{i}\stackrel{(1)}{\Psi}
  \right.
  \nonumber\\
  && \quad\quad\quad
  \left.
    +  8 H^{3} \stackrel{(1)}{\Psi} \stackrel{(1)}{\nu}_{i}
    +  4 H^{2} \stackrel{(1)}{\Psi} \partial_{\eta}\stackrel{(1)}{\nu}_{i}
    -  4 H \partial_{\eta}H \stackrel{(1)}{\Psi} \stackrel{(1)}{\nu}_{i}
  \right.
  \nonumber\\
  && \quad\quad\quad
  \left.
    +  4 H^{2} \partial_{\eta}\stackrel{(1)}{\Psi} \stackrel{(1)}{\nu}_{i}
  \right]
  \label{eq:calFXi-i-eta-def}
  ,
\end{eqnarray}
\begin{eqnarray}
  {}_{{\cal F}}\!\Xi_{ij}
  &:=&
   \frac{2 a^{2}}{H^{2}} \left(
    - 3 D_{i}\stackrel{(1)}{\Psi} D_{j}\stackrel{(1)}{\Psi}
    - 2 H^{2} \gamma_{ij} \left(\stackrel{(1)}{\Psi}\right)^{2}
    -   H \gamma_{ij} \stackrel{(1)}{\Psi} \partial_{\eta}\stackrel{(1)}{\Psi}
  \right.
  \nonumber\\
  && \quad\quad\quad
  \left.
    + 2 H D_{i}\stackrel{(1)}{\Psi} \stackrel{(1)}{\nu}_{j}
    + 2 H D_{j}\stackrel{(1)}{\Psi} \stackrel{(1)}{\nu}_{i}
  \right.
  \nonumber\\
  && \quad\quad\quad
  \left.
    +   H \stackrel{(1)}{\Psi} \partial_{\eta}\stackrel{(1)}{\chi}_{ij}
    + 2 H^{2} \stackrel{(1)}{\Psi} \stackrel{(1)}{\chi}_{ij}
  \right)
  \label{eq:second-flat-Poisson-gauge-trans-i-j}
  .
\end{eqnarray}
Together with
Eqs.~(\ref{eq:final-flat-gauge-second-metric-eta-eta})--(\ref{eq:final-flat-gauge-second-metric-i-j}),
we obtain the components of
Eq.~(\ref{eq:L-ab-in-gauge-X-def-second-flat-gauge}) as follows:
\begin{eqnarray}
  - 2 a^{2} \wt{\phi_{2\fg}}
  &=&
  - 2 a^{2} \stackrel{(2)}{\Phi}
  + 2 \partial_{\eta}{}_{{\cal F}}Y_{\eta}
  - 2 H {}_{{\cal F}}Y_{\eta}
  + {}_{{\cal F}}\!\Xi_{\eta\eta}
  ,
  \label{eq:second-flat-Poisson-gauge-trans-eta-eta-2}
  \\
  a^2(\eta)\left(
    \widetilde{B_{2\fg}}_{,i} - \widetilde{S_{2\fg i}}
  \right)
  &=&
    a^{2} \stackrel{(2)}{\nu}_{i}
  + \partial_{\eta}{}_{{\cal F}}Y_{i} 
  + D_{i}{}_{{\cal F}}Y_{\eta}
  - 2 H {}_{{\cal F}}Y_{i}
  + {}_{{\cal F}}\!\Xi_{i\eta}
  \label{eq:second-flat-Poisson-gauge-trans-eta-i-2}
  ,
  \\
  a^2(\eta) \widetilde{h_{2\fg ij}}
  &=&
  - 2 a^{2} \stackrel{(2)}{\Psi} \gamma_{ij} 
  + a^{2} \stackrel{(2)}{\chi}_{ij}
  \nonumber\\
  &&
  + D_{i}{}_{{\cal F}}Y_{j}
  + D_{j}{}_{{\cal F}}Y_{i}
  - 2 H \gamma_{ij} {}_{{\cal F}}Y_{\eta}
  + {}_{{\cal F}}\!\Xi_{ij}
  \label{eq:second-flat-Poisson-gauge-trans-i-j-2}
  .
\end{eqnarray}

%****************************************************************

The trace part of
Eq.~(\ref{eq:second-flat-Poisson-gauge-trans-i-j-2}) is given by 
\begin{eqnarray}
  0 &=&
  - 6 a^{2} \stackrel{(2)}{\Psi}
  + 2 D^{k}{}_{{\cal F}}Y_{k}
  - 6 H {}_{{\cal F}}Y_{\eta}
  + \gamma^{ij}{}_{{\cal F}}\!\Xi_{ij}
  \label{eq:second-flat-Poisson-gauge-trans-i-j-2-trace}
\end{eqnarray}
and the traceless part is 
\begin{eqnarray}
  a^2(\eta) \widetilde{h_{2\fg ij}}
  &=&
    a^{2} \stackrel{(2)}{\chi}_{ij}
  + D_{i}{}_{{\cal F}}Y_{j}
  + D_{j}{}_{{\cal F}}Y_{i}
  - \frac{2}{3} \gamma_{ij} D^{k}{}_{{\cal F}}Y_{k}
  \nonumber\\
  &&
  + {}_{{\cal F}}\!\Xi_{ij}
  - \frac{1}{3} \gamma_{ij} \gamma^{kl}{}_{{\cal F}}\!\Xi_{kl}
  ,
  \label{eq:second-flat-Poisson-gauge-trans-i-j-2-traceless}
\end{eqnarray}
where ${}_{{\cal F}}Y_{j}$ is decomposed as
\begin{eqnarray}
  {}_{{\cal F}}Y_{j}
  =:
  D_{j}{}_{{\cal F}}Y_{(L)} + {}_{{\cal F}}Y_{(V)j}, \quad
  D^{j}{}_{{\cal F}}Y_{(V)j} = 0.
\end{eqnarray}
Eq.~(\ref{eq:second-flat-Poisson-gauge-trans-i-j-2-traceless}) is
then
\begin{eqnarray}
  a^2(\eta) \widetilde{h_{2\fg ij}}
  &=&
  + a^{2} \stackrel{(2)}{\chi}_{ij}
  + 2 \left(
    D_{i}D_{j}
    - \frac{1}{3} \gamma_{ij} \Delta
  \right) {}_{{\cal F}}Y_{(L)}
  + 2 D_{(i}{}_{{\cal F}}Y_{(V)j)}
  \nonumber\\
  &&
  + {}_{{\cal F}}\!\Xi_{ij}
  - \frac{1}{3} \gamma_{ij} \gamma^{kl}{}_{{\cal F}}\!\Xi_{kl}
  ,
  \label{eq:second-flat-Poisson-gauge-trans-i-j-2-traceless-2}
\end{eqnarray}
where we have used the fact that $D_{i}$ is the covariant
derivative associated with the flat metric.
Taking the divergence of
Eq.~(\ref{eq:second-flat-Poisson-gauge-trans-i-j-2-traceless-2}),
we obtain 
\begin{eqnarray}
  \frac{4}{3} D_{j}\Delta {}_{{\cal F}}Y_{(L)}
  + \Delta{}_{{\cal F}}Y_{(V)j}
  + D^{i}\left(
    {}_{{\cal F}}\!\Xi_{ij}
    - \frac{1}{3} \gamma_{ij} \gamma^{kl}{}_{{\cal F}}\!\Xi_{kl}
  \right)
  = 0\,,
  \label{eq:second-flat-Poisson-gauge-trans-i-j-2-traceless-longi}
\end{eqnarray}
and further taking the divergence of
Eq.~(\ref{eq:second-flat-Poisson-gauge-trans-i-j-2-traceless-longi}),
gives
\begin{eqnarray}
  {}_{{\cal F}}Y_{(L)}
  =
  - \frac{3}{4} \Delta^{-2}D^{j}D^{i}\left(
    {}_{{\cal F}}\!\Xi_{ij}
    - \frac{1}{3} \gamma_{ij} \gamma^{kl}{}_{{\cal F}}\!\Xi_{kl}
  \right)
  \label{eq:second-flat-Poisson-gauge-trans-i-j-2-traceless-longi-scalar-sol}
  .
\end{eqnarray}
Substituting
Eq.~(\ref{eq:second-flat-Poisson-gauge-trans-i-j-2-traceless-longi-scalar-sol})
into
Eq.~(\ref{eq:second-flat-Poisson-gauge-trans-i-j-2-traceless-longi}),
we obtain 
\begin{eqnarray}
  {}_{{\cal F}}Y_{(V)j}
  &=&
  \Delta^{-1}\left[
    D_{j}\Delta^{-1}D^{k}D^{l}
    - \gamma_{j}^{\;\;l}D^{k}
  \right]
  \left[
    {}_{{\cal F}}\!\Xi_{kl}
    - \frac{1}{3} \gamma_{kl} \gamma^{mn}{}_{{\cal F}}\!\Xi_{mn}
  \right]
  .
  \label{eq:second-flat-Poisson-gauge-trans-i-j-2-traceless-longi-vector-sol}
\end{eqnarray}
and hence we have
\begin{eqnarray}
  {}_{{\cal F}}Y_{j}
  &=&
  \Delta^{-1}
  \left[
    \frac{1}{4} D_{j}\Delta^{-1}D^{k}D^{l}
    - \gamma_{j}^{\;\;l}D^{k}
  \right]
  \left[
    {}_{{\cal F}}\!\Xi_{kl}
    - \frac{1}{3} \gamma_{kl} \gamma^{mn}{}_{{\cal F}}\!\Xi_{mn}
  \right]
  .
  \label{eq:second-flat-Poisson-calFYj-final}
\end{eqnarray}

%****************************************************************

Since
\begin{eqnarray}
  D^{k}{}_{{\cal F}}Y_{k}
  &=&
  - \frac{3}{4} \Delta^{-1}D^{l}D^{k}
  \left[
    {}_{{\cal F}}\!\Xi_{kl}
    - \frac{1}{3} \gamma_{kl} \gamma^{mn}{}_{{\cal F}}\!\Xi_{mn}
  \right]
  \label{eq:second-flat-Poisson-calFYj-div-final}
  ,
\end{eqnarray}
Eq.~(\ref{eq:second-flat-Poisson-gauge-trans-i-j-2-traceless})
yields 
\begin{eqnarray}
  \widetilde{h_{2\fg ij}}
  &=&
  \stackrel{(2)}{\chi}_{ij}
  \nonumber\\
  &&
  + \frac{1}{2a^{2}} \left[
    \Delta^{-1}\left(
        D_{i}D_{j}\Delta^{-1}D^{k}D^{l}
      - 4 \gamma_{(i}^{\;\;l} D_{j)}D^{k}
      + \gamma_{ij} D^{l}D^{k}
    \right)
    + 2\gamma_{i}^{\;\;k}\gamma_{j}^{\;\;l}
  \right]
  \nonumber\\
  && \quad\quad
  \times
  \left[
    {}_{{\cal F}}\!\Xi_{kl}
    - \frac{1}{3} \gamma_{kl} \gamma^{mn}{}_{{\cal F}}\!\Xi_{mn}
  \right]
  \label{eq:second-flat-Poisson-h2ij-relation}
  .
\end{eqnarray}
On the other hand, the trace part of
Eq.~(\ref{eq:second-flat-Poisson-gauge-trans-i-j-2-trace}) gives
\begin{eqnarray}
  {}_{{\cal F}}Y_{\eta}
  &=&
  - \frac{a^{2}}{H} \stackrel{(2)}{\Psi}
  - \frac{1}{4H} \Delta^{-1}D^{l}D^{k} \left[
    {}_{{\cal F}}\!\Xi_{kl}
    - \frac{1}{3} \gamma_{kl} \gamma^{mn}{}_{{\cal F}}\!\Xi_{mn}
  \right]
  + \frac{1}{6H} \gamma^{ij}{}_{{\cal F}}\!\Xi_{ij}
  .
  \label{eq:second-flat-Poisson-calFYeta-final}
\end{eqnarray}
Taking the divergence of
Eq.~(\ref{eq:second-flat-Poisson-gauge-trans-eta-i-2}) and
substituting Eqs.~(\ref{eq:second-flat-Poisson-calFYj-final}),
(\ref{eq:second-flat-Poisson-calFYj-div-final}), and
(\ref{eq:second-flat-Poisson-calFYeta-final}),  we obtain
\begin{eqnarray}
  \widetilde{B_{2\fg}}
  &=&
  - \frac{1}{H} \stackrel{(2)}{\Psi}
  - \frac{1}{4a^{2}}\left[
    3 \left(
      \partial_{\eta} - 2 H
    \right) \Delta^{-1}
    + \frac{1}{H}
  \right]
  \Delta^{-1}D^{l}D^{k}
  \left[
    {}_{{\cal F}}\!\Xi_{kl}
    - \frac{1}{3} \gamma_{kl} \gamma^{mn}{}_{{\cal F}}\!\Xi_{mn}
  \right]
  \nonumber\\
  &&
  + \frac{1}{6a^{2}H} \gamma^{ij}{}_{{\cal F}}\!\Xi_{ij}
  + \frac{1}{a^{2}} \Delta^{-1} D^{i}{}_{{\cal F}}\!\Xi_{i\eta}
  \label{eq:second-flat-Poisson-B2fig-relation}
  .
\end{eqnarray}
and substituting Eq.~(\ref{eq:second-flat-Poisson-B2fig-relation})
into Eq.~(\ref{eq:second-flat-Poisson-gauge-trans-eta-i-2}) gives
\begin{eqnarray}
  \widetilde{S_{2\fg i}}
  &=&
  - \stackrel{(2)}{\nu}_{i}
  \nonumber\\
  &&
  - \frac{1}{a^{2}} \left(
    \partial_{\eta} - 2 H
  \right) \Delta^{-1} \left[
    D_{i}\Delta^{-1}D^{l}D^{k} - \gamma_{i}^{\;\;l}D^{k}
  \right]
  \left[
    {}_{{\cal F}}\!\Xi_{kl}
    - \frac{1}{3} \gamma_{kl} \gamma^{mn}{}_{{\cal F}}\!\Xi_{mn}
  \right]
  \nonumber\\
  &&
  - \frac{1}{a^{2}} \left[
    {}_{{\cal F}}\!\Xi_{i\eta}
    - D_{i}\Delta^{-1}D^{k}{}_{{\cal F}}\!\Xi_{k\eta}
  \right]
  .
  \label{eq:second-flat-Poisson-S2figi-relation}
\end{eqnarray}
Finally, through
Eq.~(\ref{eq:second-flat-Poisson-gauge-trans-eta-eta-2}), we
obtain 
\begin{eqnarray}
  \wt{\phi_{2\fg}}
  &=&
  \stackrel{(2)}{\Phi}
  + \frac{1}{a^{2}} \left(
    \partial_{\eta} - H 
  \right) \left(
    \frac{a^{2}}{H} \stackrel{(2)}{\Psi}
  \right)
  \nonumber\\
  &&
  - \frac{1}{a^{2}} \left(
    \partial_{\eta} - H 
  \right) \left(
    - \frac{1}{4H} \Delta^{-1}D^{l}D^{k} \left[
      {}_{{\cal F}}\!\Xi_{kl}
      - \frac{1}{3} \gamma_{kl} \gamma^{mn}{}_{{\cal F}}\!\Xi_{mn}
    \right]
    + \frac{1}{6H} \gamma^{ij}{}_{{\cal F}}\!\Xi_{ij}
  \right)
  \nonumber\\
  &&
  - \frac{1}{2a^{2}} {}_{{\cal F}}\!\Xi_{\eta\eta}
  \label{eq:second-flat-Poisson-phi2fig-relation}
  .
\end{eqnarray}
As at linear order, Eqs.~(\ref{eq:second-flat-Poisson-h2ij-relation}),
(\ref{eq:second-flat-Poisson-B2fig-relation}),
(\ref{eq:second-flat-Poisson-S2figi-relation}), and
(\ref{eq:second-flat-Poisson-phi2fig-relation}) give the
relationship between between the variables in uniform curvature gauge and the
Poisson gauge.

%****************************************************************

Before concluding this section, we note the non-uniqueness of
the gauge invariant variables defined in
Eqs.~(\ref{eq:linear-metric-decomp}) and
(\ref{eq:L-ab-in-gauge-X-def-second-1}) as discussed at the end
of \S\ref{sec:second_Nakamura}.
In this section, we regard the choice of gauge as the specification of
the vector fields $X^{a}$ and $Y^{a}$, which are the gauge-variant
part of the first and second order metric perturbations, respectively.
On the other hand, we may also regard gauge-fixing as the
specification of the gauge invariant parts ${\cal H}_{ab}$ and
${\cal L}_{ab}$ of the metric 
perturbations in the sense discussed at the end of
\S\ref{sec:second_Nakamura}.
Actually, we may use ${}_{{\cal F}}\!X_{a}$, whose components
are given by Eqs~(\ref{eq:calFXi-choice}) and
(\ref{eq:calFXeta-choice}), as the vector field $Z_{a}$ in 
Eq.~(\ref{eq:non-uniqueness-of-gauge-invariant-variables-first}),
since the components of ${}_{{\cal F}}\!X_{a}$ are specified by
the gauge invariant variables. 
In this case, the first order metric perturbation $h_{ab}$ is
decomposed into gauge invariant part ${\cal K}_{ab}$ and
gauge-variant part $\tilde{X}^{a}$ as
Eq.~(\ref{eq:non-uniqueness-of-gauge-invariant-variables-first}),
where ${\cal K}_{ab}$ and $\tilde{X}^{a}$ are given by   
\begin{eqnarray}
  {\cal K}_{ab} = {\cal H}_{ab} + {\pounds}_{{}_{{\cal F}}\!X}g_{ab}
  , \quad
  \tilde{X}^{a} = X^{a} - {}_{{\cal F}}\!X^{a}.
\end{eqnarray}
Further components of ${\cal K}_{ab}={}_{{\cal F}}\!h_{ab}$ are
given by Eqs.~(\ref{eq:hetaetaflat})--(\ref{eq:hijflat}).
${\cal K}_{ab}$ is regarded as the realisation of the
gauge invariant variables for the first-order metric
perturbation associated with the flat gauge.

%****************************************************************

Furthermore, in the case of the second-order metric perturbation
$l_{ab}$, we define 
\begin{eqnarray}
  \label{eq:calJ-def-flat-Poisson}
  {\cal J}_{ab}
  :=
  {\cal L}_{ab}
  + 2 {\pounds}_{{}_{{\cal F}}\!X} {}_{{\cal F}}\!h_{ab}
  + \left(
      {\pounds}_{{}_{{\cal F}}\!Y}
    - {\pounds}_{{}_{{\cal F}}\!X}^{2} 
  \right)
  g_{ab}
\end{eqnarray}
and Eq.~(\ref{eq:L-ab-in-gauge-X-def-second-flat-gauge}) is
given by 
\begin{eqnarray}
  l_{ab}
  =
  {\cal J}_{ab}
  + 2 {\pounds}_{\tilde{X}}h_{ab}
  +   \left(
    {\pounds}_{Y-{}_{{\cal F}}\!Y+[X,\tilde{X}]}
    -   {\pounds}_{\tilde{X}}^{2}
  \right)g_{ab}
  .
\end{eqnarray}
Then, choosing 
\begin{eqnarray}
  \tilde{Y}^{a} = Y^{a} - {}_{{\cal F}}\!Y^{a} + [X,\tilde{X}]^{a},
\end{eqnarray}
the second-order metric perturbation $l_{ab}$ is given by 
\begin{eqnarray}
  l_{ab}
  =
  {\cal J}_{ab}
  + 2 {\pounds}_{\tilde{X}}h_{ab}
  +   \left(
    {\pounds}_{\tilde{Y}}
    -   {\pounds}_{\tilde{X}}^{2}
  \right)g_{ab}
  .
\end{eqnarray}
The tensor ${\cal J}_{ab}$ is clearly gauge invariant and
$\tilde{Y}^{a}$ satisfy the gauge transformation rule
\begin{eqnarray}
  {}_{{\cal Y}}\!\tilde{Y}^{a}
  -
  {}_{{\cal X}}\!\tilde{Y}^{a}
  &=&
  \xi_{(2)}^{a}
  + [\xi_{(1)},{}_{{\cal X}}\!\tilde{X}]^{a}
\end{eqnarray}
under the transformation 
$\Phi={\cal X}^{-1}\circ{\cal Y}$, i.e., $\tilde{Y}^{a}$ satisfy
the property (\ref{eq:gauge-trans-of-calL-and-Y}) of the
gauge-variant part of the second order metric perturbation. 
The components of the gauge invariant part 
${\cal J}_{ab}$ are given by
Eqs.~(\ref{eq:final-flat-gauge-second-metric-eta-eta})--(\ref{eq:final-flat-gauge-second-metric-i-j}) and so
${\cal J}_{ab}$ is regarded as the realisation of the
gauge invariant variables for the second order metric 
perturbation associated with the flat gauge.

%*******************************************************************

At the end of
\S\ref{sec:second_standard}
we noted that the second order tensor perturbation is gauge dependent.
Therefore, the expression for the gauge invariant second order tensor perturbation
will differ depending on the choice of gauge. 
This is due to the fact that in perturbation theory beyond linear
order, in which mode-couplings occur due to the non-linearity of the
system, the notion of the transverse-traceless part of the metric
perturbation cannot be identified uniquely with the gravitational
waves. As shown for the second order case in Section \ref{sec:second_standard},
 the
gauge invariant higher order transverse-traceless perturbation has contributions
from the second order tensor perturbation and other first order metric
potentials.  In other words, we might say that the second order
gravitational waves are also generated from the the first order
gravitational potentials as discussed in literature ~\cite{tensors}.

%*******************************************************************

%%%%%%%%%%%%%%%%%%%%%%%%%%%%%%%%%%%%%%%%%%%%%%%%%%%%%%%%%%%%%%%%%%%%%%
\section{Summary and discussions}
\label{sec:Summary_and_discussions}
%%%%%%%%%%%%%%%%%%%%%%%%%%%%%%%%%%%%%%%%%%%%%%%%%%%%%%%%%%%%%%%%%%%%%%

%*******************************************************************

In this paper, we have compared and contrasted two different
approaches to metric based cosmological perturbation theory, and
have derived the relationship between the standard approach \'a
la Bardeen~\cite{Bardeen-1980}, which was discussed in
MW2009~\cite{Malik:2008im}, and the formalism studied in
KN2010~\cite{Nakamura:2010yg}.
We started by introducing the basics of relativistic
perturbation theory, and then presented first- and second-order
cosmological perturbation theory in the MW2009 and KN2010
approach, respectively.
Finally, in \S\ref{sec:Comparison_of_different_formulations}, we
compared the two formalisms directly for both the longitudinal
or Poisson gauge and the uniform curvature gauge.

%*******************************************************************

Both approaches recognise the spurious gauge artefacts present
in perturbation theory, and adopt different techniques in order
to remove them.
In the MW2009 approach, one perturbs the metric tensor and then
inspects the gauge transformations of the metric perturbations,
using these to eliminate the gauge freedom and hence construct
gauge invariant variables.
The KN2010 approach splits the perturbation to the metric into a
gauge invariant and gauge variant part, and then writes all
equations in terms of these gauge invariant variables.
In the MW2009 approach, the gauge choice is made by
specifying the gauge generating vector, $\xi^\mu$.
On the other hand, in the KN2010 approach the gauge is defined
through the gauge variant vector $X^a$ (or, at second order,
$Y^a$). 
We showed that the gauge invariant variables that are used in
KN2010 are equivalent to the usual gauge invariant variables of
the Poisson gauge (so that the Poisson gauge is specified
through $X^a=0=Y^a$).
When relating KN2010 formalism to the MW2009 approach in the
uniform curvature gauge, we simply obtain the usual relationship
between the gauge invariant variables in the Poisson gauge and
those in the uniform curvature gauge.
Thus, we have shown that the two approaches are equivalent.

%*******************************************************************

While this result is not necessarily surprising, since both the
approaches are based upon metric cosmological perturbation
theory, showing this equivalence is a good consistency check for
the KN2010 approach.
Furthermore, there may be certain problems for which it could be
advantageous to use one approach over the other.
Having shown this equivalence, and having a working knowledge of
both theories, may enable one to more easily solve the problem
in hand.

%*******************************************************************

%%%%%%%%%%%%%%%%%%%%%%%%%%%%%%%%%%%%%%%%%%%%%%%%%%%%%%%%%%%%%%%%%%%%%%
\section*{Acknowledgements}
%%%%%%%%%%%%%%%%%%%%%%%%%%%%%%%%%%%%%%%%%%%%%%%%%%%%%%%%%%%%%%%%%%%%%%
A.J.C, K.A.M, and K.N are grateful to the participants and organisers of the
long-term workshop ``GC2010'' (YITP-T-10-01) where this project began.
In particular, they thank Misao Sasaki and
Takahiro Tanaka for their hospitality during the visit. 
A.J.C is supported by the Science
and Technology Facilities Council (STFC) and K.A.M is
supported, in part, by STFC under Grant ST/G002150/1.
The computer algebra package {\sc Cadabra} ~\cite{Cadabra} was used to obtain
some of the gauge transformations in Section \ref{sec:second_standard}.

%****************************************************************

%%%%%%%%%%%%%%%%%%%%%%%%%%%%%%%%%%%%%%%%%%%%%%%%%%%%%%%%%%%%%%%%%%%%%%
\appendix
%%%%%%%%%%%%%%%%%%%%%%%%%%%%%%%%%%%%%%%%%%%%%%%%%%%%%%%%%%%%%%%%%%%%%%

%*********************************************************************
\section*{References}

%%%%%%%%%%%%%%%%%%%%%%%%%%%%%%%%%%%%%%%%%%%%%%%%%%%%%%%%%%%%%
%%%%%%%%%%%%%%%%%%%%%%%%%%%%%%%%%%%%%%%%%%%%%%%%%%%%%%%%%%%%%

\end{document}